\documentclass[fleqn,usenatbib]{mnras}

\usepackage{newtxtext,newtxmath}

\usepackage[T1]{fontenc}
\usepackage{ae,aecompl}

\usepackage{graphicx}	
\usepackage{amsmath}	
\usepackage{amssymb}	

\newbox\grsign \setbox\grsign=\hbox{$>$} \newdimen\grdimen \grdimen=\ht\grsign
\newbox\simlessbox \newbox\simgreatbox
\setbox\simgreatbox=\hbox{\raise.5ex\hbox{$>$}\llap
     {\lower.5ex\hbox{$\sim$}}}\ht1=\grdimen\dp1=0pt
\setbox\simlessbox=\hbox{\raise.5ex\hbox{$<$}\llap
     {\lower.5ex\hbox{$\sim$}}}\ht2=\grdimen\dp2=0pt

\def\simless{\mathrel{\copy\simlessbox}}
\newbox\simppropto
\setbox\simppropto=\hbox{\raise.5ex\hbox{$\sim$}\llap
     {\lower.5ex\hbox{$\propto$}}}\ht2=\grdimen\dp2=0pt

\title[A Deep View of a Fossil Relic in the Galactic  Bulge]{A Deep View of a Fossil Relic in the Galactic Bulge: The Globular Cluster HP\,1 \thanks{Based on observations obtained at the Gemini Observatory, which is operated by the Association of Universities for Research in Astronomy, Inc., under a cooperative agreement with the NSF on behalf of the Gemini partnership: the National Science Foundation (United States), National Research Council (Canada), CONICYT (Chile), Ministerio de Ciencia, Tecnolog\'{i}a e Innovaci\'{o}n Productiva (Argentina), Minist\'{e}rio da Ci\^{e}ncia, Tecnologia e Inova\c{c}\~{a}o (Brazil), and Korea Astronomy and Space Science Institute (Republic of Korea).}\thanks{Based on observations made with the NASA/ESA Hubble Space Telescope, obtained from the data archive at the Space Telescope Science Institute. STScI is operated by the Association of Universities for Research in Astronomy, Inc. under NASA contract NAS 5-26555.}}

\author[Kerber et al.]{L. O. Kerber,$^{1,2}$\thanks{E-mail: lokerber@uesc.br}
M. Libralato,$^{3,4,5}$
S. O. Souza,$^{1}$
R. A. P. Oliveira,$^{1}$
S. Ortolani,$^{4,5}$
\newauthor{
A. P\'erez-Villegas,$^{1}$
B. Barbuy,$^{1}$
B. Dias,$^{6,7}$ 
E. Bica,$^{8}$
D. Nardiello$^{4,5}$
}
\\
$^{1}$Universidade de S\~ao Paulo, IAG, Rua do Mat\~ao 1226,
Cidade Universit\'aria, S\~ao Paulo 05508-900, Brazil\\
$^{2}$ Universidade Estadual de Santa Cruz, Rodovia Jorge Amado km 16, 
Ilh\'eus 45662-000, Bahia, Brazil\\
$^{3}$ Space Telescope Science Institute, 3700 San Martin Drive,
Baltimore, MD 21218, USA  \\
$^{4}$Dipartimento di Fisica e Astronomia, Universit\`a di Padova, I-35122 Padova,
 Italy\\
$^{5}$INAF-Osservatorio Astronomico di Padova, Vicolo dell'Osservatorio 5,
I-35122 Padova, Italy\\
$^{6}$European Southern Observatory, Alonso de Cordova 3107, Santiago, Chile\\
$^{7}$Departamento de F\'\i sica, Facultad de Ciencias Exactas, Universidad Andr\'es Bello, Av. Fernandez Concha 700, Las Condes, Santiago, Chile\\
$^{8}$Universidade Federal do Rio Grande do Sul, Departamento de Astronomia,
CP 15051, Porto Alegre 91501-970, Brazil\\
}

\date{Accepted 2018 December 10. Received 2018 November 14; in original form 2018 May 8}

\pubyear{2015}

\begin{document}
\label{firstpage}
\pagerange{\pageref{firstpage}--\pageref{lastpage}}
\maketitle

\begin{abstract}
HP\,1 is an $\alpha$-enhanced and moderately metal-poor bulge globular cluster with a blue horizontal branch. These combined characteristics make it a probable relic of the early star formation in the innermost Galactic regions. Here we present a detailed analysis of a deep near-infrared (NIR) photometry of HP\,1 obtained with the NIR GSAOI+GeMS camera at the Gemini-South telescope. $J$ and $K_{\rm S}$ images were collected with an exquisite spatial resolution (FWHM $\sim 0.1$ arcsec), reaching stars at two magnitudes below the MSTO. We combine our GSAOI data with archival F606W-filter \textit{HST} ACS/WFC images to compute relative proper motions and select bona fide cluster members. Results from statistical isochrone fits in the NIR and optical-NIR colour-magnitude diagrams indicate an age of $12.8^{+0.9}_{-0.8}$ Gyr, confirming that HP\,1 is one of the oldest clusters in the Milky Way. The same fits also provide apparent distance moduli in the $K_{\rm S}$ and $V$ filters in very good agreement with the ones from 11 RR Lyrae stars. By subtracting the extinction in each filter, we recover a heliocentric distance of $6.59^{+0.17}_{-0.15}$ kpc. Furthermore, we refine the orbit of HP\,1 using this accurate distance and update and accurate radial velocities (from high resolution spectroscopy) and absolute proper motions (from Gaia DR2), reaching mean perigalactic and apogalactic distances of $\sim$0.12 and $\sim$3 kpc respectively.
\end{abstract}





\section{Introduction}



Galactic globular clusters (GCs) play a key role in the understanding of the formation history of the Milky Way (MW) because they are excellent tracers of its early chemical evolution, going from the outer halo to the innermost Galactic regions. In particular, a very interesting class of objects are the bulge GCs possessing moderately metal-poor ([Fe/H]$\lesssim -1.0$) and $\alpha$-enhanced ([$\alpha$/Fe]$\gtrsim +0.3$) stars, with a blue horizontal branch (BHB). The relatively high metallicity of these GCs is due to an early fast chemical enrichment in the central parts of the MW  as shown by \citet[][]{Cescutti+08}. These objects might be relics of an early generation of long-lived stars formed in the proto-Galaxy. A dozen of them were identified by \citet{Barbuy+98, Barbuy+09}, and for this reason we have been performing high-resolution spectroscopy studies \citep{Barbuy+06, Barbuy+07, Barbuy+09, Barbuy+14, Barbuy+16, Barbuy+18}, as well as optical and near-infrared (NIR) multi-epoch photometry to build proper-motion (PM) cleaned colour-magnitude diagrams (CMDs) \citep[e.g.][]{Ortolani+11, Rossi+15}. Recently, this hypothesis has been tested for two bulge GCs, NGC\,6522 and NGC\,6626, employing a statistical isochrone fitting method on deep and decontaminated \textit{Hubble Space Telescope} (\textit{HST} optical CMDs, confirming that these clusters are old with a range of ages between $\sim $12.5 and 13.0 Gyr \citep{Kerber+18}.

Age derivation for a larger sample of bulge GCs in order to test the initial hypothesis concerning their very old ages is
however made difficult, due to the severe extinction, strong field contamination and relatively large distance. The bulge GCs represent therefore a  last frontier in terms of observational challenge to study such objects in the Galaxy. 
Furthermore, the total mass of a bulge GC in only a few cases surpasses $\sim 10^{5} M_{\odot}$, making the disentanglement between cluster and field exclusively possible with proper-motion cleaning techniques. 
In fact, even with the exquisite spatial resolution and accurate photometry from the \textit{HST}, it is not an easy task to build deep proper-motion-cleaned CMDs in the UV/visual for bulge GCs. 
Not by chance, the two most comprehensive \textit{HST} surveys of GCs -- the Advanced Camera for Surveys (ACS) survey of GCs \citep{Sarajedini+07} and the \textit{HST} UV Legacy Survey of GCs \citep{Piotto+15} -- were focused on halo and disk objects, almost neglecting those in the bulge.  

NIR is the natural solution to go deep in the stellar content of the bulge GCs. Using longer wavelengths than in the visible, the extinction is dramatically reduced (e.g, $A_{K_{\rm{S}}} \sim 0.12 A_{V}$) and the adaptive optics (AO) is significantly more efficient, reaching almost the telescope diffraction limit (full-width at half maximum, FWHM, $\sim 0.1$ arcsec).

Boosted by the continuing improvement of the third generation of infrared detectors based on HgCdTe technology in the last two decades, significant efforts have been done to collect NIR photometry of bulge GCs. Notably, projects using ground-based 2-4m class telescopes from the European Southern Observatory (ESO) \citep[e.g.,][]{Valenti+07, Valenti+10}, particularly the \textit{Vista Variable in the Via L\'actea} (VVV) Survey \citep[e.g.,][]{Minniti+10, Minniti+17, Cohen+17}, have expanded our knowledge about these objects by covering almost all stars in the red giant branch (RGB) and horizontal branch (HB) in different filters. 
In combination with several spectroscopy studies, in particular those based on high-resolution spectroscopy 
of individual red giants 
in the visible \citep[e.g.,][]{Barbuy+14, Gratton+15, Munoz+17, Villanova+17} and in the NIR \citep[e.g.,][]{Lee+04, Origlia+Rich04, Valenti+11, Schiavon+17}, important results to establish a more comprehensive scenario for the formation and evolution of the bulge GCs has been obtained. This includes the metallicity distribution and chemical abundances analysis.

However, the lack of deep NIR photometry reaching at least one magnitude below the main sequence turn-off (MSTO) 
of the bulge GCs has severely hampered the determination of reliable ages for these systems, therefore preventing the achievement of the age-metallicity relation (AMR) for the innermost Galactic region. Fortunately, the immediate perspective to overcome this challenge is positive thanks to the current NIR detector on board of the \textit{HST} (the NIR channel of the Wide Field Camera 3 - WFC3/IR) and the multi-conjugate adaptive optics systems (MCAO) at the large ground-based telescopes \citep[e.g.,][]{Ferraro+16}.



Recently, \citet{Cohen+18} presented deep NIR and optical-NIR CMDs for 16 bulge GCs built with \textit{HST} photometry from WFC3/IR and ACS/Wide Field Channel (WFC) images. These homogeneous and high-quality data for a significant number of clusters, which includes HP\,1, represent an important milestone for any future study concerning bulge GCs.
Furthermore, recent NIR images using MCAO systems, such as the Multi-Conjugate Adaptive Optics Demonstrator (MAD) at the Very Large Telescope (VLT), and the Gemini South Adaptive Optics Imager (GSAOI) combined with the Gemini Multi-Conjugated Adaptive Optics System (GeMS) at the Gemini South Telescope, provided unprecedented results related to GCs \citep[e.g..][]{Turri+15,Massari+16}, specifically to the ones presented in the bulge \citep{Saracino+15, Saracino+16}. 


We obtained observing time at the Gemini South Telescope 
to collect deep $J$ and $K_{\rm S}$ images of HP\,1 using the GSAOI+GeMS camera. 
HP\,1 is an inner bulge GC placed at just $3^{\rm o}.33$ from the Galactic center, with Galactic coordinates $l=-2^{\rm o}.58$, $b=+2^{\rm o}.12$ \citep{harris96}. A metallicity of [Fe/H]$=-1.06\pm0.10$ was determined by the analysis of high-resolution spectra of eight red giant stars \citep{Barbuy+06,Barbuy+16}, in good agreement with results from low-resolution spectroscopy of $\rm{[Fe/H]}=-1.17\pm0.07$ by \citep{Dias+16a}. In addition, all these studies indicated that HP\,1 is $\alpha$-enhanced, with [$\alpha$/Fe]$\sim+0.3$. 
Previously, \citet{Ortolani+11} built a $K_{\rm S}$ vs. $V-K_{\rm S}$ PM cleaned CMD for HP\,1, using two-epoch photometry separated by 14.25 years, the first (in $V$) collected in 
1994 with the NTT telescope and the second (in $J$ and $K_{\rm S}$) obtained in 2008 with the MAD at the VLT. This CMD of excellent quality, confirmed the 
presence of a BHB, but barely reached the MSTO ($K_{\rm S} \sim 17.5$). Their analysis pointed out to a very old age ($\sim$ 13.7 Gyr) and a metallicity of [Fe/H]$\sim-1.0$ (estimated from the RGB slope).

The main goal of this work is to check if HP\,1 is a fossil relic. For this purpose, we used the aforementioned GSAOI+GeMS data in combination with  \textit{HST} images obtained with the ACS/WFC in F606W filter to compute relative PMs and isolate probable cluster members to plot in NIR and optical-NIR CMDs. RR Lyrae stars presented in the OGLE Collection of Variable Stars (CVS) for the Galactic bulge \citep{Soszynski+14},\footnote{http://ogledb.astrouw.edu.pl/$\sim$ogle/CVS/} and the ($m_{\rm F606W} - K_{\rm S}$) vs. ($J-K_{\rm S}$) colour-colour diagram were used to provide additional constraints for the heliocentric distance, reddening and helium abundance.
Furthermore, we performed an orbital analysis of HP\,1 using the cluster distance derived in the present work, the recent and accurate radial velocities from high resolution spectroscopy \citep{Barbuy+16}, and the absolute PMs from the Gaia DR2 catalog \citep{Gaia18}, with the purpose of constraining the maximum galactocentric distance where these fossil relic stars could be found.

The present work is structured as follow. The Gemini and \textit{HST} data are presented in Section \ref{datared}, whereas CMDs and the relative PM computation are described in Section \ref{cmdpm}. 
In Section \ref{isochrones}, the adopted stellar evolutionary models are indicated. Section \ref{constraints} is dedicated to the colour-colour diagram and the RR Lyrae stars. Then, the statistical isochrone fitting method following a Bayesian approach and its results are reported in Sections \ref{isot_fit} and \ref{results}. A discussion about age and distance determinations is in Sect. \ref{discussion}. We also present an orbital analysis of HP\,1 in Section \ref{orbit}. Finally, conclusions and some perspectives are drawn in Section \ref{conclusions}.

\begin{figure*}
\includegraphics[width=\textwidth,angle=0]{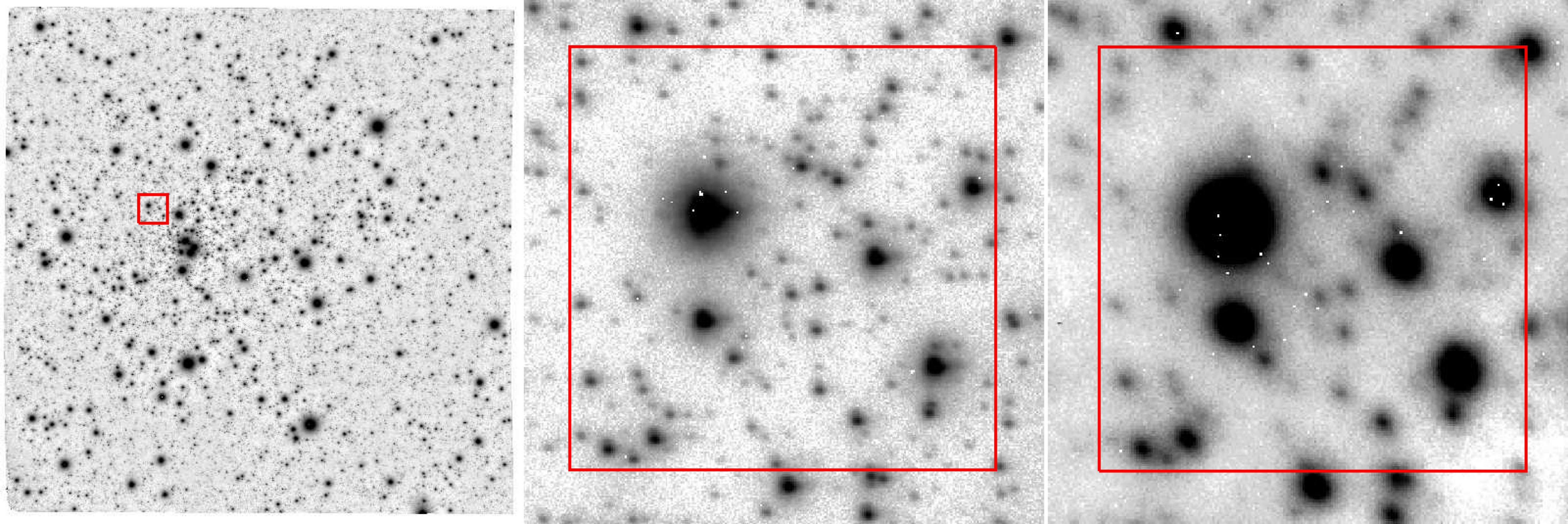}
    \caption{FoV of HP~1. The mosaic of the $K_{\rm S}$-filter GSAOI images used in this paper is shown in the left panel. The FoV covered is about $1.4 \times 1.5$ arcmin$^2$. In the middle and right panels, we present a comparison between a GSAOI@Gemini-S and MAD@VLT exposure in a $5 \times 5$ arcsec$^2$ box. Images are in logarithmic scale; North is up and East to the left. }
    \label{HP1_MAD_vs_GSAOI}
\end{figure*}

\section{Observations}\label{datared}

\subsection{GSAOI$+$GeMS@Gemini-South}\label{gsaoired}

GSAOI is a mosaic of four Rockwell Hawaii-2RG 2048$\times$2048
pixel$^2$ detectors that cover a field of view (FoV) of about
85$\times$85 arcsec$^2$ on the sky. The pixel scale of the detectors
is 20 mas pixel$^{-1}$.

The GSAOI observations of HP\,1 analyzed in this work were obtained in May 4, 2017 as part of the program GS-2017A-Q44 (PI: L. Kerber). 
The field was observed in $J$ and $K_{\rm S}$ filters with 14 images of exposure time of 30\,s (DIT$\times$NDIT $=$ 30\,s$\times$1) per filter. Five additional $K_{\rm S}$-filter images with an exposure time of 10\,s were achieved to extend the analysis to bright stars otherwise saturated in longer exposures.
Figure \ref{HP1_MAD_vs_GSAOI} presents the mosaic of GSAOI+GeMS $K_{\rm S}$ images for HP\,1 used in this work, and a comparison between a GSAOI@Gemini-South and a MAD@VLT image. 

The median value of the FWHM for $J$ and
$K_{\rm S}$ images is about 0.15 and 0.10 arcsec, respectively. 
The FWHM of $J$ images is overall more stable than that of $K_{\rm S}$ exposures, which for the latter filter can be as large as 0.15 arcsec in the rightmost edges of chips 1 and 4 (see also Appendix~\ref{GD}). 
As such, the quality of the data varies across the FoV. Furthermore, $J$-filter images are shallower than those in $K_{\rm S}$, and limited our investigation of the MS of HP\,1 to about one magnitude below the MSTO.

The reduction of GSAOI data was performed as in
\citet{Libralato+14,Libralato+15}, following the
procedures and softwares described in \citet{Anderson+06}. We
refer to these papers for the detailed description of the data
reduction. Here we provide a brief overview and discuss in detail the
differences with respect to the reference papers.

For each chip, we created a hot-pixel mask following the prescriptions
given by the official GSAOI
website\footnote{\href{https://www.gemini.edu/sciops/instruments/gsaoi/}{https://www.gemini.edu/sciops/instruments/gsaoi/}}. We
adopted the most-recent dark frames available in the archive with an
exposure time as close as possible to that of the scientific
exposures. Hot-pixel structures change over time, therefore our
hot-pixel masks represent an average snapshot of the hot-pixel
distribution in our scientific images.

Then, we constructed a master flat-field frame for each chip and
filter. Flat-field exposures can be obtained with (``DOME FLAT ON'')
or without (``DOME FLAT OFF'') flat-field light. The latter images are
useful to estimate the amount of thermal emission of the telescope in
$H$ and $K_{\rm S}$ observations. The master flat-field images were
obtained by computing the normalized $\sigma$-clipped median value of
each pixel of the ``DOME FLAT ON'' flat-field exposures. For $K_{\rm
  S}$-filter frames, we subtracted in each flat-field image the median
value of the $K_{\rm S}$-filter ``DOME FLAT OFF'' frames before
combining them in a master flat-field image. In addition, flat-field
exposures allowed us to create a map of bad-pixels in the detector.

We corrected scientific images by means of these master flat-field
frames. We did not subtract any dark frame from our images because of
the negligible amount of dark current in GSAOI detectors. Then, we
flagged hot, bad and saturated pixels not to use them in the
subsequent analysis. Saturation was set to 32\,000 counts. Finally, we
computed the median sky values in a $5 \times 5$ grid and subtracted
the sky contribution in a given pixel of the detector according to
this table.

We constructed a $5 \times 5$ array of empirical, spatially-varying
point-spread functions (PSFs) for each chip/exposure/filter based
on the available bright and relatively isolated stars
available in each image. PSFs were computed with the software
\texttt{img2psf\_GSAOI}, a modified version of the original
\texttt{img2psf\_WFI} made by \citet{Anderson+06}. These
empirical PSFs were used to measure position and flux of all
detectable sources in the field. As a diagnostic parameter for the PSF
fit, we defined the ``quality-of-PSF-fit'' (\texttt{QFIT}) as the
absolute fractional error in the PSF fit to the star
\citep{Anderson+08}. The \texttt{QFIT} value ranges between 0
and 1: the closer to 0, the better the PSF fit. Stellar positions were
corrected for geometric distortion (GD). We refer to Appendix~\ref{GD}
for the detailed description of the GD solution of GSAOI detectors.

For each filter, we cross-identified the same stars in the meta
catalogs (see Appendix~\ref{GD}) to create a common reference frame
with average positions and fluxes of all sources. The final GSAOI
catalog contains only stars measured in at least three $K_{\rm S}$-
and $J$-filter exposures. Photometry was calibrated to the Two Micron All Sky Survey
\citep[2MASS,][]{Skrutskie+06} photometric system by using the
MAD catalog of \citet{Ortolani+11}.

\subsection{\textit{HST}}\label{hstred}

We made use of \textit{HST} data
obtained in August 2016 with the ACS/WFC camera under GO-14074 (PI: Cohen; see \citet{Cohen+18}) to
compute relative PMs. The data set consists in four long
(495\,s) and one short (40\,s) F606W-filter exposures.
Near-IR images obtained with the IR channel of the Wide-Field Camera 3 (WFC3) are
also available for this cluster. However, we chose not to use them
because of the worse astrometric precision of WFC3/IR data with
respect to that of ACS/WFC.

We used ACS/WFC \texttt{\_flc} exposures, \texttt{\_flt} images
pipeline corrected for charge-transfer-efficiency (CTE) defects
\citep[see][]{Anderson+Bedin10}. Positions and fluxes of all stars
in each image were extracted with a single finding wave by fitting
empirical, spatially- and time-varying PSFs obtained by tailoring the
publicly-available\footnote{\href{http://www.stsci.edu/~jayander/STDPSFs/}{http://www.stsci.edu/$\sim$jayander/STDPSFs/}.}
library PSFs for each ACS/WFC F606W-filter exposure. Stellar positions
were corrected for geometric distortion (GD) by using the GD solution provided by
\citet{Anderson+King06}. Finally, as for GSAOI data, we computed a master frame
catalog by means of six-parameter linear transformations. The ACS/WFC
F606W photometry was registered on the Vega mag system as in, e.g.,
\citet{Bellini+17a}.

\section{Proper-motion-cleaned CMDs}\label{cmdpm}

\begin{figure*}
  \includegraphics[width=\textwidth,angle=0]{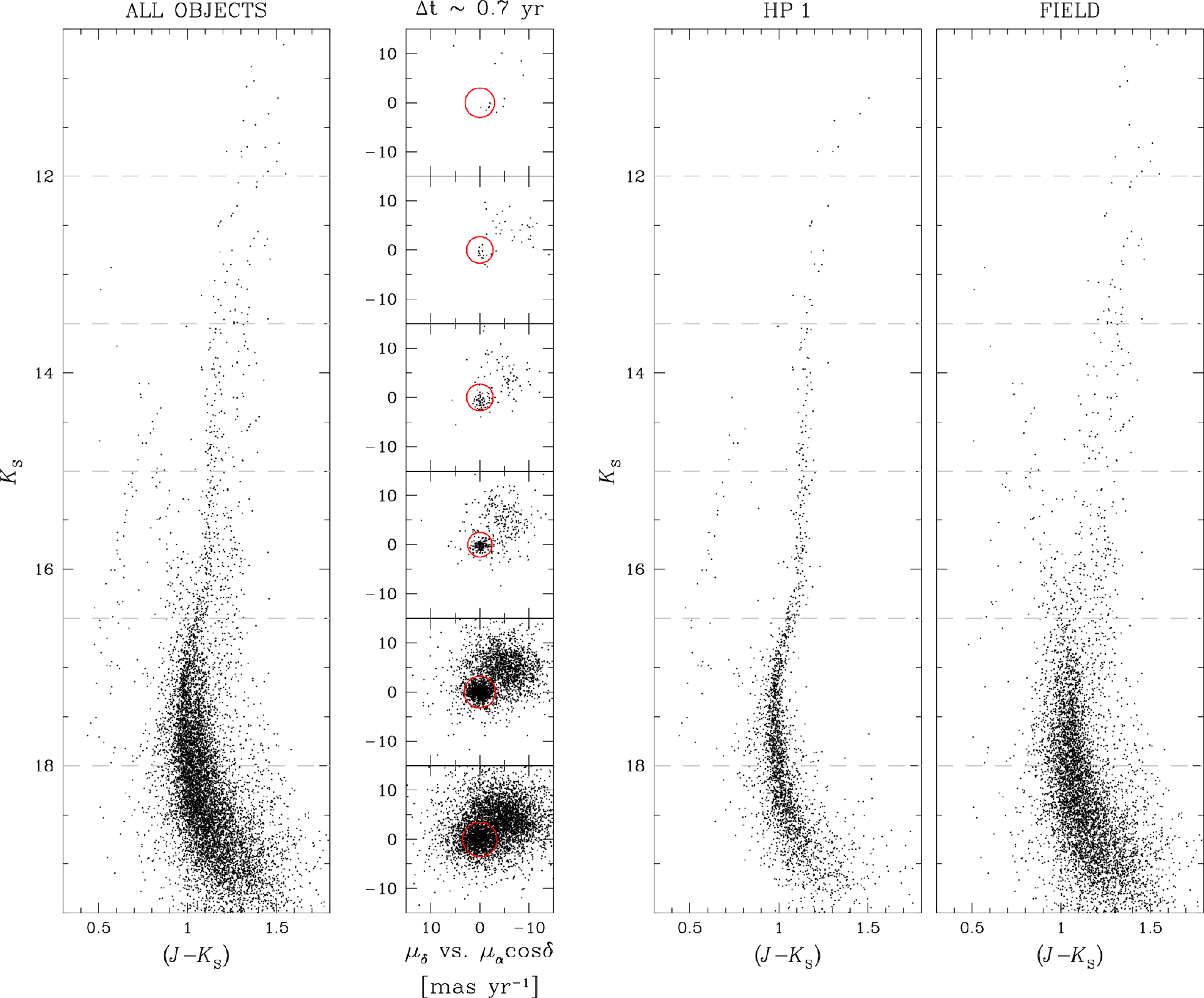}
  \caption{In the left panel, we present the $K_{\rm S}$
    vs. $(J-K_{\rm S})$ CMD of all stars in the FoV. Only
    well-measured (both astrometrically and photometrically) stars are
    shown. We divided our sample in six 1.5-magnitude bins to better
    infer the cluster membership. The gray, dashed horizontal lines
    are the limits of each magnitude bin. In the middle-left VPDs, the
    red circles are centered at (0,0) mas yr$^{-1}$ and have a radius
    of 3.0, 2.7, 2.7, 2.5, 3.2, 3.5 mas yr$^{-1}$ from top to bottom,
    respectively. Stars within these circles are considered as likely
    members of HP\,1. The CMD of the cluster members is presented in
    the middle-right panel. In the CMD on the right we plot only field
    stars. The separation between HP\,1 and field objects is not
    clear, especially at faint magnitudes, and might result in a
    contamination of sources members of one population into the sample
    of the other.}
  \label{cmd1}
\end{figure*}

Field contamination in the CMD of HP\,1 represents a major limitation
for the analysis of this bulge GC, in particular along the
MS. Therefore, we computed the relative PMs of the objects in the FoV
by combining the GSAOI and \textit{HST} catalogs, which offer a
temporal baseline of 0.71 yr.

We applied local transformations to compute relative PMs similarly to
\citet[see also \citealt{Anderson+06}]{Massari+13}. 
For each star in our GSAOI catalog, we
selected a sample of bright, well-measured, close-by likely cluster
stars to compute the coefficients of the six-parameter
linear transformations (rigid shifts in the two coordinates, one
rotation, one change of scale, one term that represent the deviation
from orthogonality between the two axes, and one term that accounts
for the change of the relative scale between the two axes) to
transform the position of the target star, as measured in the
\textit{HST} catalog to the GSAOI reference frame. The relative
displacement of a given star is computed as the difference between its
GSAOI-based and the \textit{HST}-based transformed positions on the
GSAOI reference-frame system. The relative PM errors are estimated as the sum
in quadrature of the positional uncertainty in the GSAOI and
\textit{HST} catalogs (in mas) divided by the temporal baseline. We
computed the PMs in  relative terms. As such, the distribution
of HP\,1 stars is centered on the origin of the vector-point diagram
(VPD), while field stars lie in different locations according to their
relative motion with respect to HP\,1. 

We found colour-related and spatial systematic effects in our final relative PMs,
therefore we corrected them similarly to \citet{Anderson+06}
and \citet{Bellini+18}, respectively. Finally, we used the Gaia Data Release 1 \citep{Gaia16a,Gaia16b} catalog to transform
our relative PMs from GSAOI pixels year$^{-1}$ into $\mu_{\alpha} \cos\delta$ and
$\mu_\delta$ relative PMs in mas yr$^{-1}$.

In Figure~\ref{cmd1}, we show the $K_{\rm S}$ vs. $(J-K_{\rm S})$ CMDs
and VPDs. In this and all other figures of the paper, CMDs are
corrected for the effects of the differential reddening as described
in \citet{Bellini+17b}, which is grounded on the methodology
described in \citet{Milone+12}. We refer to these papers for
the detailed description of the correction. Furthermore, in all CMDs
we present only well-measured stars, defined as having: (i) $x$ and $y$ positional rms, (ii) $K_{\rm S}$ and $J$ magnitude rms,  (iii) $K_{\rm S}$ and $J$ \texttt{QFIT}, and (iv) 1-D relative PM error smaller than the 75$^{\rm th}$-percentile value at any given magnitude. Most of the rejected stars are at the faint end of the MS and in regions imaged in chips 1 and 4 because of the worse quality of the data (see Sect.~\ref{gsaoired}).

The left panel of Figure~\ref{cmd1} shows the CMD of all objects, while
the two rightmost panels show likely cluster members and field
objects, respectively. The VPDs illustrate the membership-selection
criteria we adopted to identify bona-fide cluster members. We
divided our sample in six magnitude bins of 1.5 magnitudes each. In
each bin, we drew a circle centered in the origin of the VPD (the mean
motion of HP\,1, by construction). All stars enclosed by the circles
are considered members of HP\,1. The selection radius adopted in each
magnitude bin was chosen as a compromise between including field stars
with a motion similar to that of the cluster, and excluding HP\,1
members with a large relative PM. Furthermore, the radius of the circle in the
VPDs is larger for saturated and very-faint stars to take into account
for their larger relative PM errors.

\begin{figure}
	\centering
  \includegraphics[width=0.7\columnwidth,angle=0]{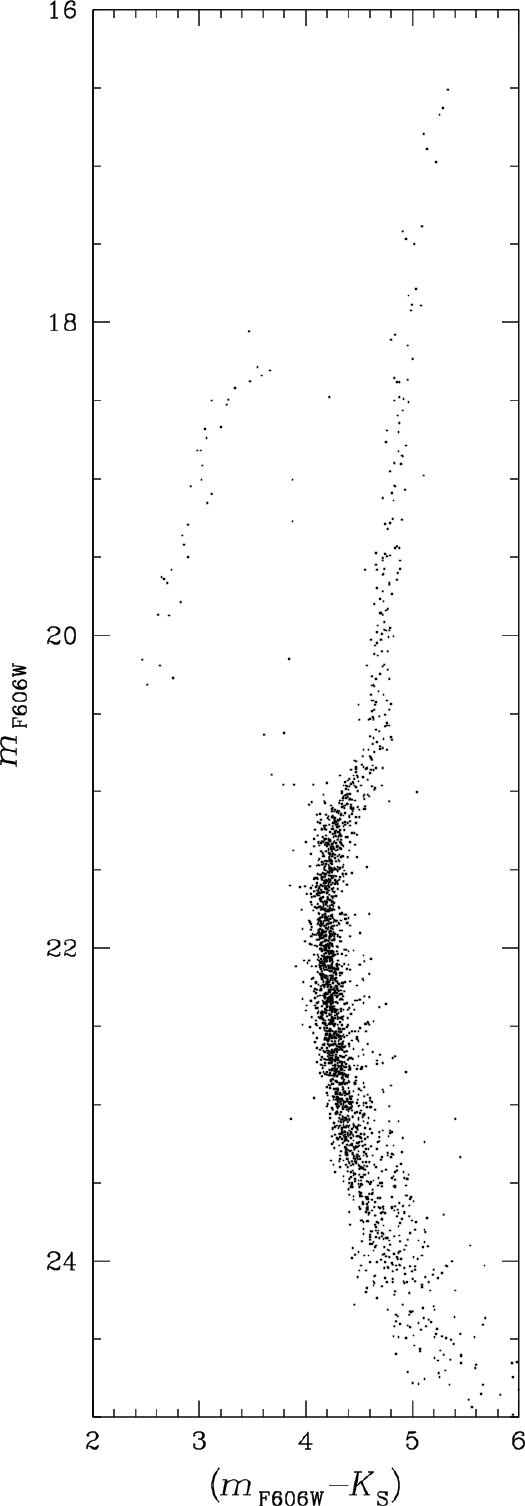}
  \caption{$m_{\rm F606W}$ vs. $(m_{\rm F606W}-K_{\rm S})$ CMD of likely
  members of HP\,1.}
  \label{cmd2}
\end{figure}

Finally, Figure~\ref{cmd2} presents the $m_{\rm F606W}$ vs. $(m_{\rm F606W}-K_{\rm S})$ CMD for stars members of HP~1 according to their relative PMs.


\subsection{Non-linear effects in GSAOI photometry}

As previously reported by \cite{Massari+16} and
\citet{Saracino+16}, GSAOI+GeMS photometry presents
non-linearity effects.  These issues mainly affect RGB stars,
steepening the slope of the RGB.

We adopted an a-posteriori, empirical correction to mitigate the
non-linearity effects.  We compared the $K_{\rm S}$ photometry of the
stars in the long exposures with that obtained from the short
exposures zero-pointed to the long-exposure photometric system.
For stars at the MSTO, for which we do not expect non-linearity
effects, we found an agreement between the long- and the short-based
magnitudes.  In the magnitude interval within which the linearity
effects becomes sizeable in the long but not in the short exposures
($13 < K_{\rm S} < 15$), we measured stars systematically fainter in
the long than in the short exposures.  For brighter
magnitudes, the non-linearity effects and saturation problems are also
present in the short exposures, thus limiting any further
investigation. As such, we adopted only stars with $13 < K_{\rm S} <
15$ to compute our correction.

\begin{figure}
  \includegraphics[width=\columnwidth,angle=0]{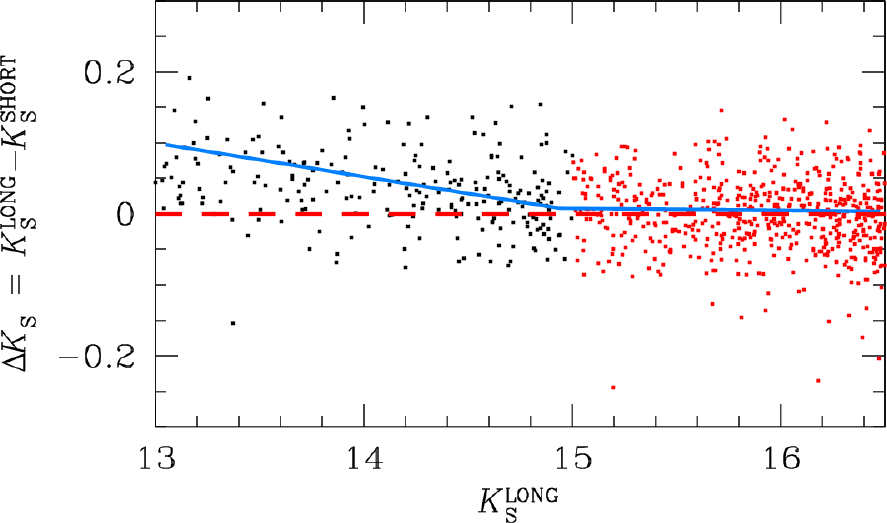}
  \caption{Difference between long- and short-exposure $K_{\rm S}$ magnitudes.
  Red points refer to stars for which the non-linearity effects are negligible
  in both long and short exposures, while black points mark objects for which
  the effects are sizeable only in the long exposures. The blue line represents
  the linear relation made by hand to interpolate (and correct) the non-linearity
  in the GSAOI@Gemini-S photometry.
  }
  \label{deltaks}
\end{figure}

We found that a straight line is enough to model the non-linearity
effects in the magnitude range $13 < K_{\rm S} < 15$ (Figure~\ref{deltaks}). 
The correction to add to the GSAOI $K_{\rm S}$ magnitudes brighter than
$K_{\rm S} < 15$ was extrapolated by this straight
line. We do not have $J$-filter short exposures, therefore we corrected
the $J$ photometry using the same linear relation as for the $K_{\rm
  S}$ magnitudes.

\begin{figure}
  \includegraphics[width=\columnwidth,angle=0]{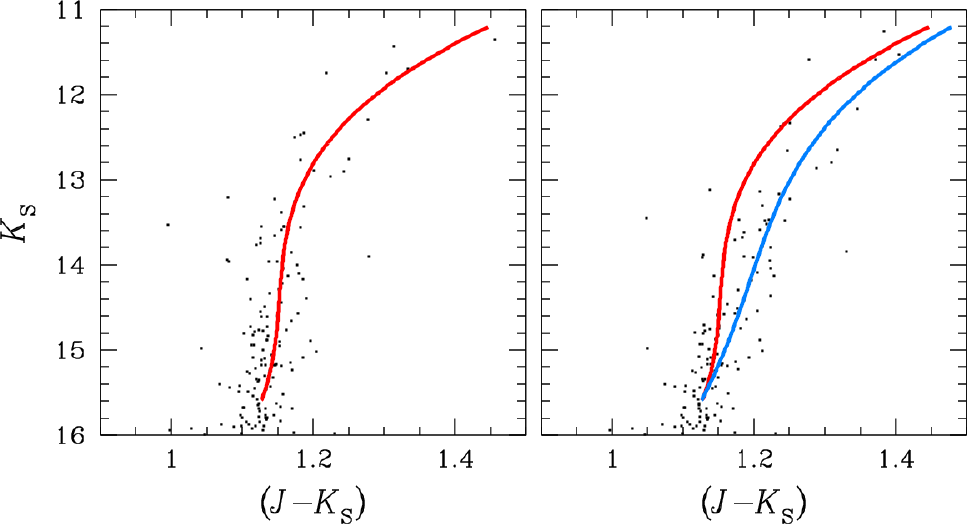}
  \caption{Difference between the original (left) and
    non-linearity-corrected (right) $K_{\rm S}$ vs. $(J-K_{\rm S})$
    CMDs of HP\,1. The red and the blue lines represent the HP\,1 RGB
    fiducial lines in the original and corrected photometry,
    respectively.}
  \label{cmd3}
\end{figure}

The original (left panel) and non-linearity-corrected (right panel) CMDs
are shown in Figure~\ref{cmd3}.  We drew a fiducial line of the RGB
stars using the original (red line) and the corrected (blue line) photometry to
highlight the difference in the RGB slope before and after the correction.
The correction is fairly good up to $K_{\rm S} \sim 13$, while for brighter
stars it is underestimated. Although this correction is enough for a more reliable isochrone fit, the correction is applied only to improve the cosmetics of the CMDs. In the following sections, we limited our analysis to stars fainter than $K_{\rm S}=15$.
Hereafter,
we used only the corrected photometry in our CMDs unless explicitly declared
otherwise.


\section{Stellar Evolutionary Models}
\label{isochrones}


To compare our GSAOI and \textit{HST} data with theoretical stellar evolutionary models, we employed two sets of isochrones:
\begin{itemize}
\item \textit{Dartmouth Stellar Evolutionary Database} \citep[DSED;][]{Dotter+08}\footnote{\href{http://stellar.dartmouth.edu/models/}{http://stellar.dartmouth.edu/models/}} and
\item \textit{A Bag of Stellar Tracks and Isochrones} \citep[BaSTI][]{Pietrinferni+06}\footnote{\href{http://albione.oa-teramo.inaf.it}{http://albione.oa-teramo.inaf.it}}.
\end{itemize}

In both cases we retrieved $\alpha$-enhanced ([$\alpha$/Fe]=+0.40) isochrones with canonical helium ($Y\sim0.25$), metallicities similar to the one from HP\,1 ([Fe/H]$\sim -1.0$), and covering ages from 10.0 to 15.0 Gyr in steps of 0.20 Gyr. DSED and BaSTI isochrones are available in the ACS/WFC@\textit{HST} photometric system, but only the former were converted to the 2MASS photometric system by the original developers.   
On the other hand, BaSTI NIR colours are originally in the Johnson-Cousins-Glass photometric system, so they were first converted to the \citet{Bessel+Brett88} system and then to the 
2MASS photometric system using the transformations presented in the final 2MASS data release.\footnote{\href{http://www.astro.caltech.edu/~jmc/2mass/v3/transformations/}{http://www.astro.caltech.edu/~jmc/2mass/v3/transformations}}

\begin{figure*}
    \includegraphics[width=\columnwidth]{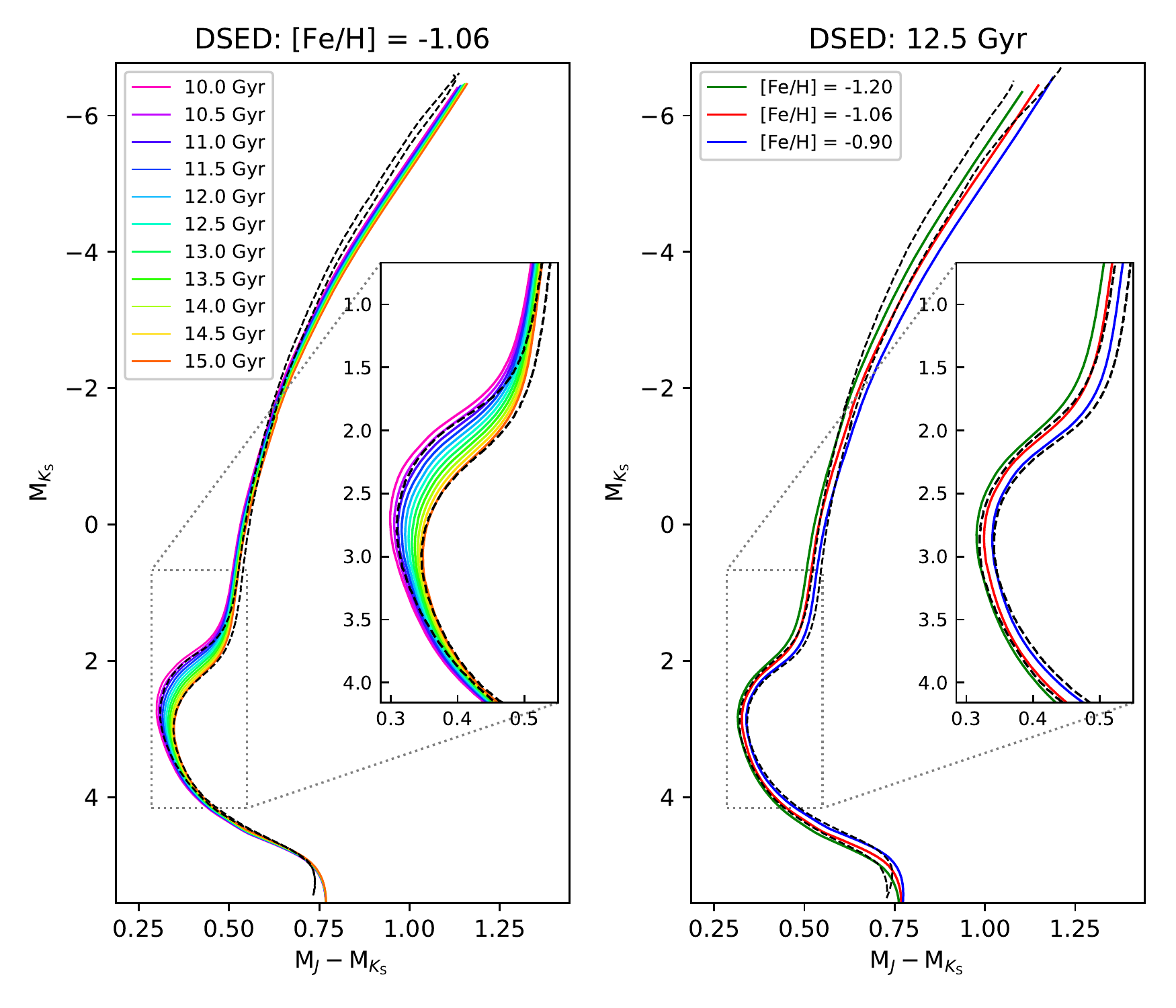}
    \includegraphics[width=\columnwidth]{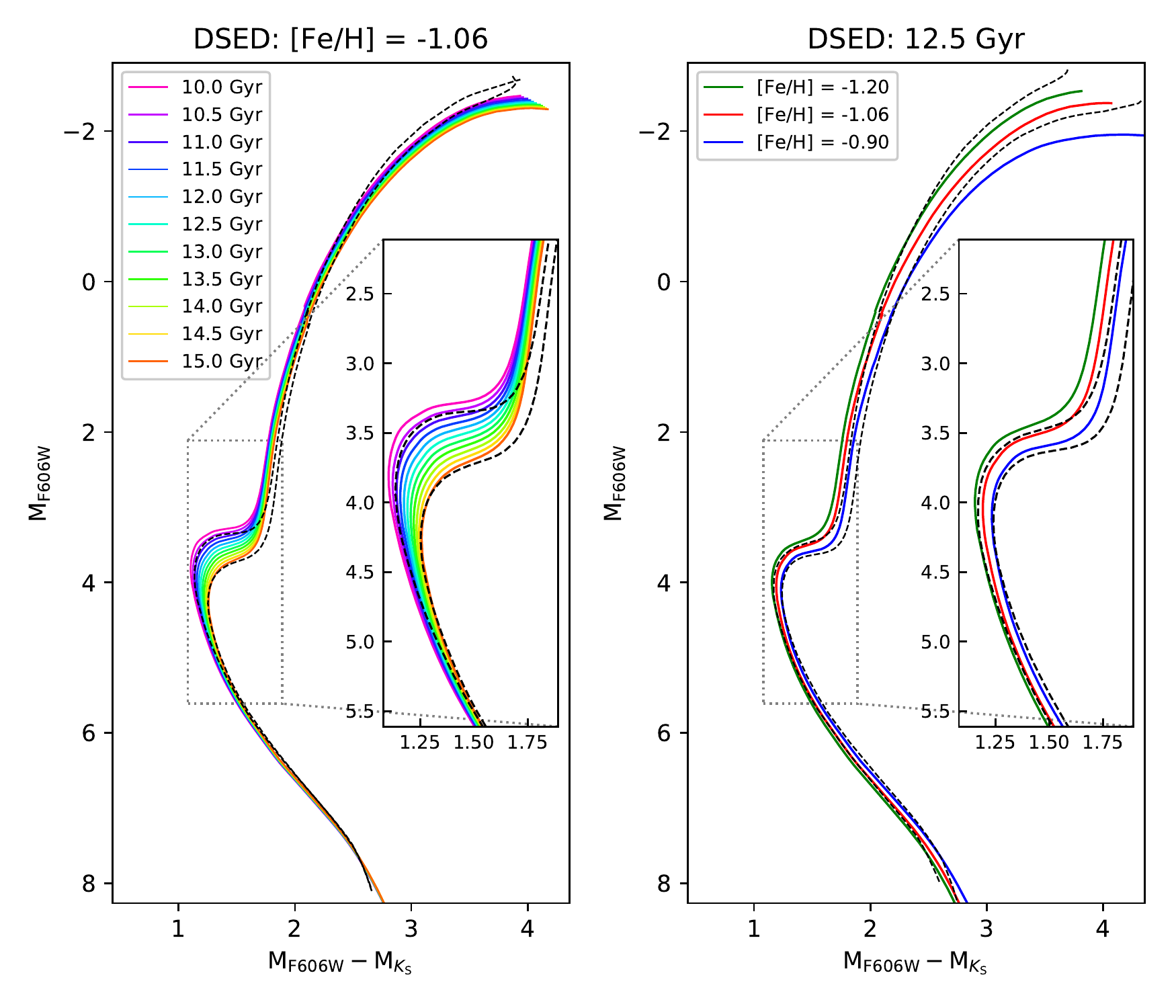}
    \caption{Isochrones from DSED models showing the effect of age (10.0 to 15.0 Gyr, for a metallicity [Fe/H]$=-1.06$) and metallicity ([Fe/H]$= -1.20, -1.06, -0.90$, for an age of 12.5 Gyr) in the $M_{K_{\rm S}}$ vs. $M_{J}-M_{K_{\rm S}}$ CMD (left panels) and in the $M_{\rm F606W}$ vs. $(M_{\rm F606W} - M_{K_{\rm S}})$ (right panels) CMD. All models have [$\alpha$/Fe]$=+0.40$ and canonical helium abundance ($Y\sim 0.25$). BaSTI isochrones with the extreme age and [Fe/H] values are also presented (dashed lines).}
    \label{fig_isochrones_DSED}
\end{figure*}

Figure \ref{fig_isochrones_DSED} illustrates the DSED and BaSTI isochrones used in this work, showing the effect of the age and metallicity in the $M_{K_{\rm S}}$ vs. $M_{J} - M_{K_{\rm S}}$ and $M_{\rm F606W}$ vs. $(M_{\rm F606W} - M_{K_{\rm S}})$ CMDs. Interpolations in metallicity were performed in order to compute isochrones with $-1.26 \leq$ [Fe/H] $\leq -0.86$ in steps of 0.02 dex, therefore fully covering the accurate value of [Fe/H]$=-1.06 \pm0.10$ from high-resolution spectroscopic analysis of 
\citet{Barbuy+16}.

Although the DSED and BaSTI models seem to produce isochrones that are almost indistinguishable, they present slightly distinct age scales. 
Differently from the DSED code, 
the effects of atomic diffusion were not incorporated in the BaSTI one.
This leads to an apparent overestimation in the ages from the BaSTI models of about 0.9 Gyr in relation to the models where the atomic diffusion is included \citep[][and references therein]{Cassisi+98,Cassisi+99}. 
In Section \ref{isot_fit} we address this issue in order to find not only the systematic differences in age between the BaSTI and DSED, but also those in the other parameters. 

\section{Constraints from Colour-colour diagrams and RR Lyrae stars}
\label{constraints}


Taking the advantage of deep multiband photometry, from optical (F606W) to NIR ($J$ and $K_{\rm S}$), as well as the presence of RR Lyrae stars towards the cluster that were identified by the OGLE Collection of Variable Stars \citep{Soszynski+14}, we obtained some constraints on the reddening and distance values of HP\,1. 

Due to the limitation in the photometric depth, the previous simultaneous determinations of these parameters were based on red giant stars.
Using a $V$ vs. $V-I$ colour-magnitude diagram obtained with NTT data, \citet{Ortolani+97} recovered colour excess $E(B-V)=1.19$ and a distance $d_{\odot}=6.75\pm0.60$ kpc (or an intrinsic distance modulus of $(m-M)_{0}=14.15\pm0.20$). These authors found a RGB slope similar to the one of NGC\,6752 ([Fe/H]$\sim -1.5$).
From the analysis of a $K$ vs. $(J-K)$ CMD built with NIR images collected with the same telescope, \citet{Valenti+10} determined very similar reddening ($E(B-V)=1.19$) and distance modulus ($(m-M)_{0}=14.17$, equivalent to $d_{\odot}=6.8$ kpc), but a slightly less metal-poor solution ([Fe/H]$=-1.12$), in very good agreement with the recent spectroscopic result from \citet{Barbuy+16} ([Fe/H]$=-1.06\pm0.10$). 

Finally, \citet{Ortolani+11} built a proper-motion-cleaned CMD using their previous NTT images in the $V$ filter as the first-epoch and MAD@VLT images in the $J$ and $K_{\rm S}$ filters as the second one. By means of an isochrone fit using Padova models with [Fe/H]$\sim-1.0$, they determined a colour excess $E(V-K_{\rm S})=3.3$ ($\sim E(B-V)=1.17$) and a distance of $d=7.1\pm0.5$ kpc, in both cases assuming a $R_{V}=A_{V}/E(B-V)=3.2$. They also recovered a short distance of 6.8 kpc comparing the magnitude difference in $V$ between the HP\,1 HB stars at the RR Lyrae level and that of the bulge field. In this case they assumed as the solar distance to the Galactic centre a value of 8.0 kpc. 

\subsection{Colour-colour diagram and reddening}

Colour-colour diagrams are useful tools to determine the reddening of a cluster since they do not depend on the distance. Despite this fact, they should be used with caution since colours -- in particular involving optical and NIR -- might present systematic uncertainties caused by small zero point corrections in the photometric calibrations. Furthermore, when $E(B-V)\gtrsim 0.3$ (or $A_{V}\gtrsim 1.0$), as in the present case, the reddening and extinction dependence on the effective temperature ($T_{\rm eff}$) become relevant, particularly in the visible range. It means that in this high reddening regime a single value for ratios between the extinction in a given photometric band ($A_{\lambda}$) and $A_{V}$ is no longer a good approximation, particularly to analyse very accurate photometry in the optical. Furthermore, it is important to point out that R$_{\rm V}$ depends on the colour and on the reddening value
\citep{McCall04}. For our typical CMD colours and extinction, $R_{\rm V}=3.2$ appear more suitable than the standard value of 3.1. 

\begin{figure}
	\includegraphics[width=\columnwidth]{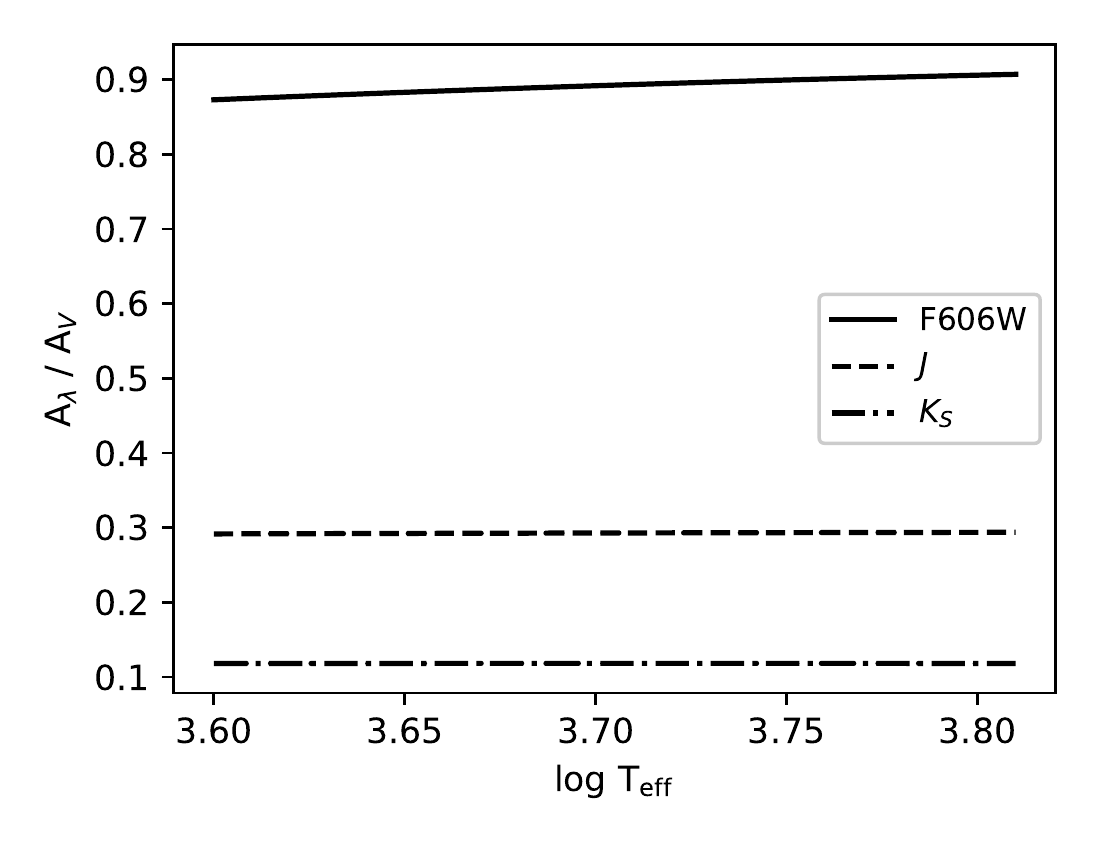}
    \caption{$A_{\rm F606W}/A_{V}$ (solid line), $A_{J}/A_{V}$ (dashed line), $A_{K_{\rm S}}/A_{V}$ (dotted-dashed line) as a function of the star effective temperature.}
    \label{fig_Alambda_vs_logTeff}
\end{figure}

To compute $A_{\lambda}/A_{V}$ as a function of $\log{T_{\rm eff}}$
we used the CMD 3.0 web interface,\footnote{\href{http://stev.oapd.inaf.it/cgi-bin/cmd} {http://stev.oapd.inaf.it/cgi-bin/cmd} }
which implemented the results from Girardi et al. (2008) models, to retrieve PARSEC isochrones \citep{Bressan+12} with $E(B-V)=0.00$ and $E(B-V)=1.15$.
Therefore, the desired relations are obtained by measuring the magnitude differences between these isochrones as a function of $\log{T_{\rm eff}}$ (Figure \ref{fig_Alambda_vs_logTeff}).
As expected, the $A_{\lambda}/A_{V}$ variation is only significant in the F606W filter ($\sim 5\%$), whereas it is almost negligible in the NIR filters. 

\begin{figure}
	\includegraphics[width=\columnwidth]{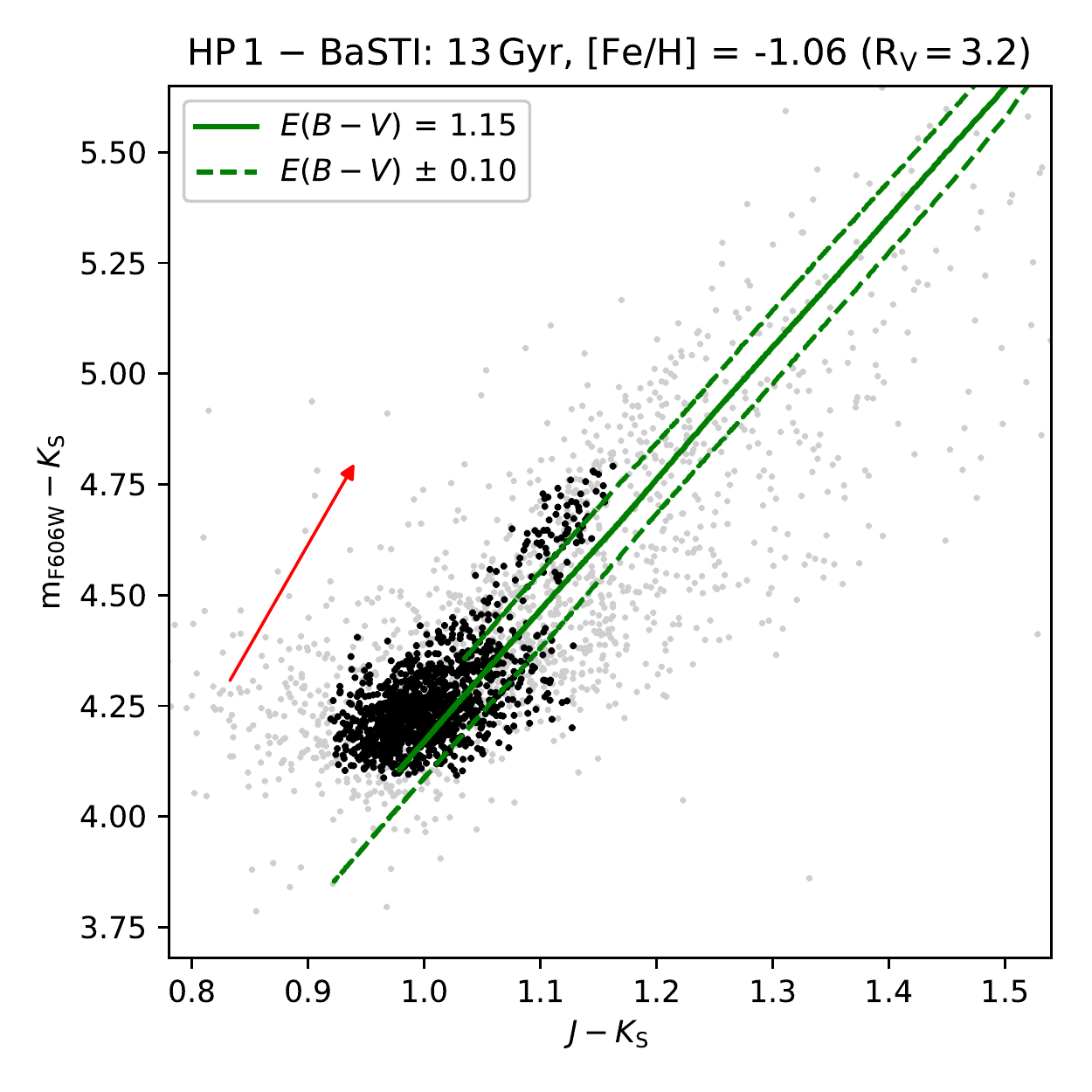}
    \caption{Colour-colour diagram of HP\,1. The green solid line correspond to a BaSTI isochrone with age of 13.0 Gyr, [Fe/H]$=-1.06$, R$_{V}=\frac{A_{V}}{E(B-V)}=3.2$ and $E(B-V)=1.15$. An uncertainty of $0.10$ in $E(B-V)$ is limited by the green dashed lines. The red arrow correspond to a reddening vector of $E(B-V)$=0.20. Black points correspond to the ones used in the isochrone fits (see Sect. \ref{isot_fit}.) }
    \label{redd1}
\end{figure}

Figure \ref{redd1} presents the colour-colour $(m_{\rm F606W}-K_{\rm S})$ vs. $(J-K_{\rm S})$ diagram for HP\,1.
An $\alpha$-enhanced ([$\alpha$/Fe]$=+0.40$]) BaSTI isochrone with [Fe/H]$=-1.06$ and 13.0 Gyr was chosen as a reference for the colours. 
A reddening value of $E(B-V)=1.15\pm0.10$ generate a good overall fit, in accordance with previous determinations from literature. 
This range in reddening 
was used by us to estimate the distances of the RR Lyrae, whereas a wide rage of $\Delta$E($B-V$)$=0.360$ was employed to impose initial limits in the isochrone fits (See Sect.\ref{isot_fit}).

\subsection{RR Lyrae and distances}
\label{RRLyrae}

\begin{figure}
	\includegraphics[width=\columnwidth,angle=0]{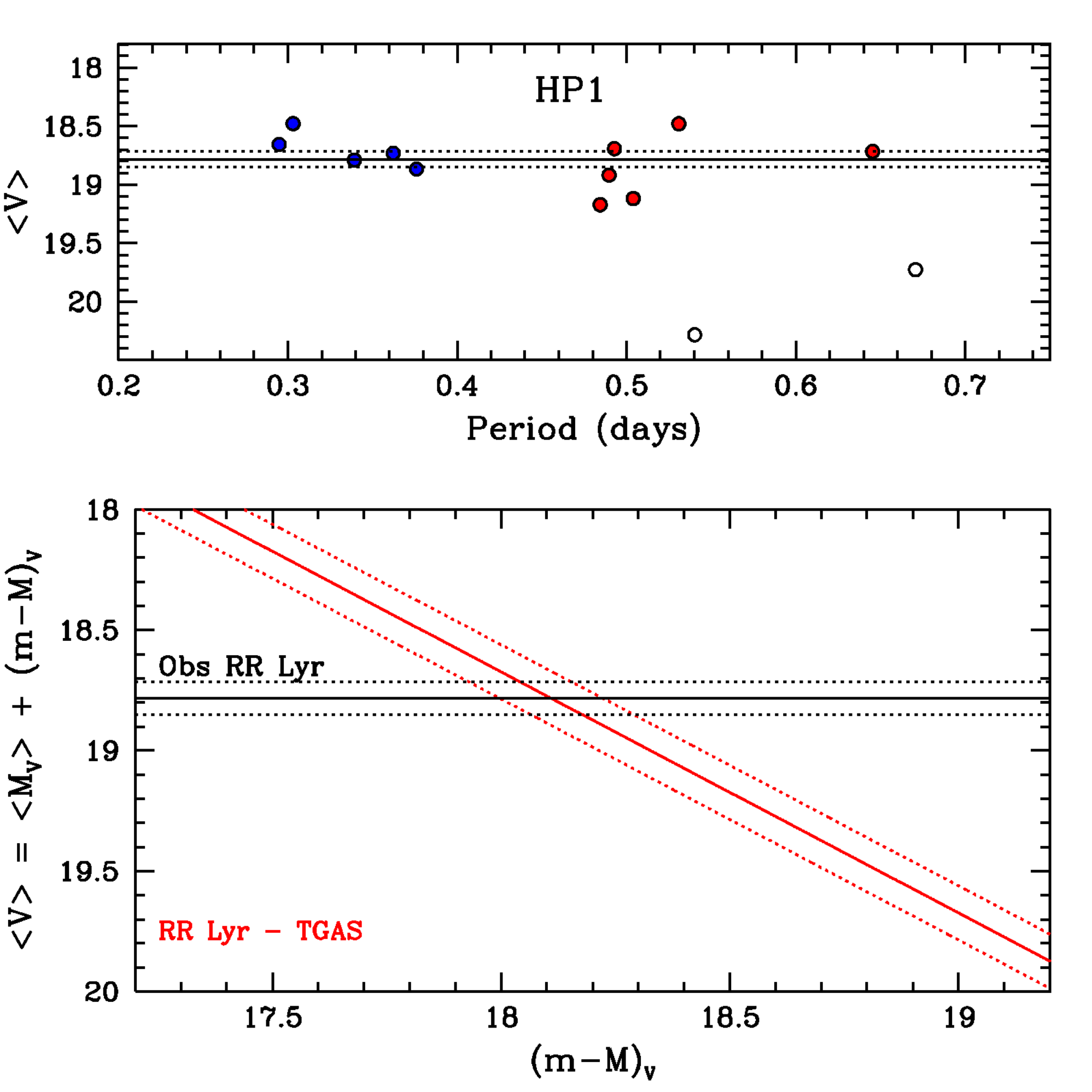}
    \caption{Mean $V$ magnitudes vs. period of the RRab (red circles) and RRc (blue circles) Lyrae stars in HP\,1 (top panel) and the expected apparent distance modulus in $V$ for these stars assuming an $M_{V}=0.67\pm0.11$ (red lines) (bottom panel). All RR Lyrae stars come from the OGLE CVS. The absolute magnitude in $V$ was determined using [Fe/H]$=-1.06\pm0.10$ and the empirical $M_{V}$-[Fe/H] relations from the \citet{Gaia17} using the TGAS. The mean of the mean V magnitudes (solid black line) and its standard deviation (dotted lines) is show in both panel. Two rejected RR Lyrae stars are also presented (open circles).
    }
    \label{RRLyr_V}
\end{figure}

RR Lyrae stars are Population II pulsating stars and one of the fundamental primary distance indicators in astronomy. 
Presenting periods under a day and light curve amplitudes from few tenths of magnitude in $K_{\rm {S}}$ to $\sim$ 1 magnitude in $V$, these core helium burning stars are very common in old globular clusters with [Fe/H]$\lesssim -1.0$ since these systems present HB stars populating the instability strip. 
Therefore, RR Lyrae stars are ideal standard candles to impose independent constraints on distance in a moderately metal-poor bulge GC like HP\,1. 
Using the OGLE CVS for the Galactic bulge \citep{Soszynski+14}
we identified 13 RR Lyrae stars within 150 arcsec from the cluster center. 
The mean V magnitude and periods for these stars are presented in Figure \ref{RRLyr_V} (upper panel), where it is possible to identify two probable foreground field stars. 
The remaining 11 RR Lyrae stars are divided in 6 of RRab type and 5 of RRc type, providing a value of $18.78\pm0.07$ for the mean of the mean $V$ magnitudes ($\langle V \rangle$). 

In order to determine the apparent distance modulus for these stars, one needs a prediction for the absolute V magnitude ($M_{V}$) for the HP\,1 metallicity ([Fe/H]$=-1.06\pm0.10)$.
For this purpose, we used the very recent empirical calibrations for the $M_{V}-$[Fe/H] relation from the \citet{Gaia17} based on Tycho-\textit{Gaia} Astrometric Solution (TGAS). 
Taking into account the uncertainties in the HP\,1 metallicity, the systematics and stochastic effects from the three different methods to perform such calibration, we obtained an average value of $\langle M_{V} \rangle =0.67\pm 0.11$.
Since, by definition, the apparent distance modulus in $V$ is the difference between $V$ and $M_{V}$, the $(m-M)_{V}$ value for HP\,1 is $18.11\pm0.18$.
This result is illustrated in Figure \ref{RRLyr_V} (bottom panel) by the intersection between the observed $\langle V \rangle$ value (black line) and the one expected for RR Lyrae stars assuming the TGAS calibration (red line; $\langle V \rangle = \langle M_{V} \rangle + (m-M)_{V}$).  
Assuming an $R_{V}=3.2\pm0.10$ and a conservative value of $E(B-V)=1.15\pm0.10$, we derive a value of $A_{V}=3.68\pm0.34$ for the extinction in $V$.
This result leads us to an intrinsic distance modulus of $(m-M)_{0}=14.43\pm0.38$, which means a large and high uncertain heliocentric distance of $7.7^{+1.5}_{-1.2}$ kpc.



A more reliable and accurate result for the apparent distance modulus and distance from RR Lyrae stars can be reached using the $K_{\rm {S}}$ filter. This is a natural and straightforward consequence of a significantly lower extinction in $K_{\rm {S}}$ than in $V$ ($A_{K_{\rm S}}/A_{V}=0.1185$), as well as period-luminosity relationships with very low scattering \citep[e.g.,][]{Beaton+16}.
To take advantage of the NIR photometry, we performed a cross-match between the RR Lyrae stars from the OGLE CVS with the stars presented in the \textit{VISTA Variables in the Via Lactea} (VVV) survey \citep{Saito+12}.\footnote{\href{https://vvvsurvey.org}{https://vvvsurvey.org}}

\begin{figure}
	\includegraphics[width=\columnwidth,angle=0]{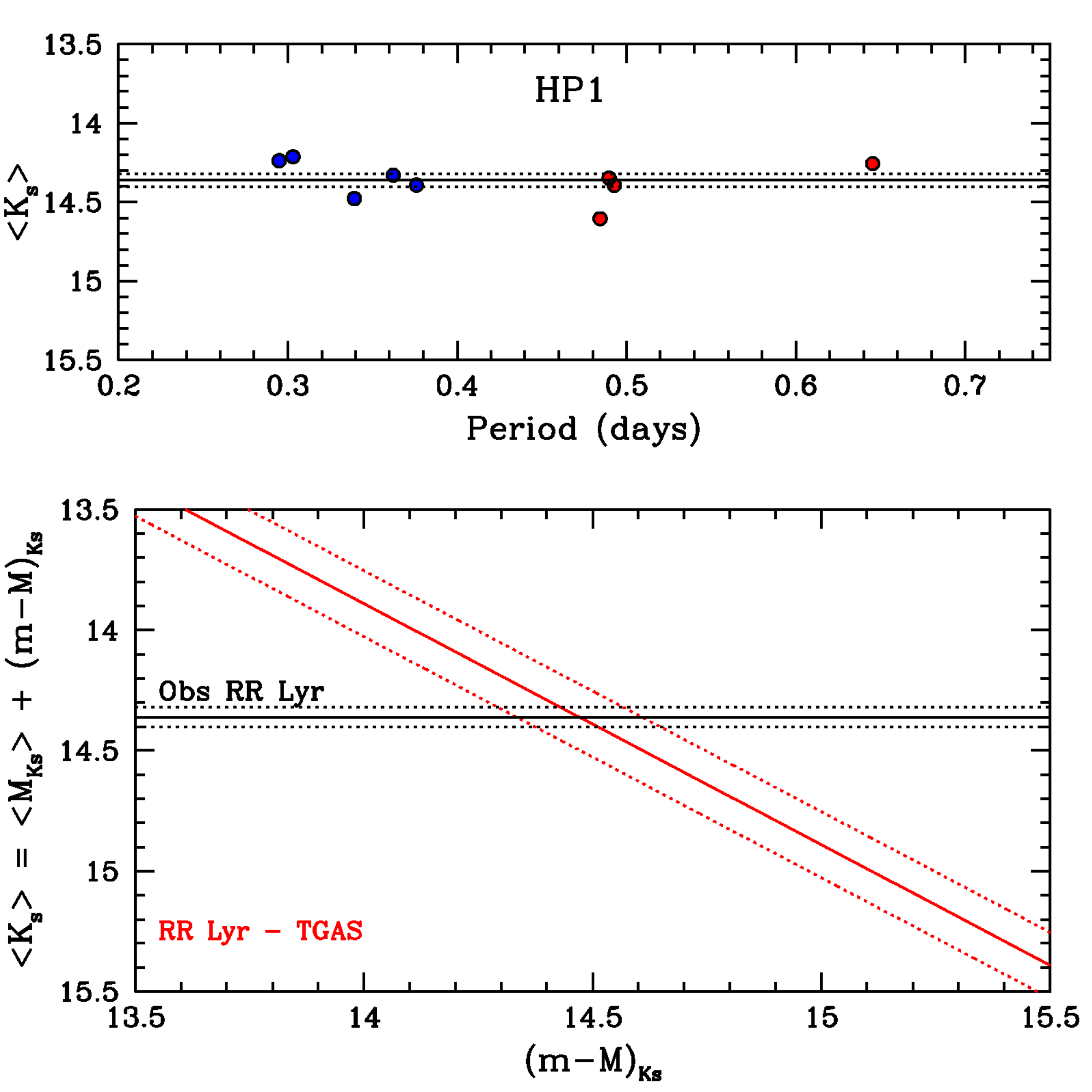}
    \caption{Same as Figure \ref{RRLyr_V}, but for the $K_{\rm S}$ filter. 
    In this case we adopted $M_{K_{\rm S}}=-0.11\pm0.14$ from the $PM_{K_{\rm S}}Z$ relations from \citet{Gaia17}. The $K_{\rm S}$ magnitudes come from the VVV survey.  
    }
    
    \label{RRLyr_Ks}
\end{figure}

Figure \ref{RRLyr_Ks} (upper panel) presents the results for the nine RR Lyrae stars recovered in this process, where a threshold of 1.0 arcsec in the source separation was applied.
The mean of the mean $K_{\rm S}$ magnitude of these stars is $14.36\pm0.04$. 
Using the mean solution of the three empirical $M_{K_{\rm S}}$--$\log{P}$--[Fe/H] relations ($PM_{K_{\rm S}}Z$) from the \citet{Gaia17}, we estimated an average absolute $K_{\rm S}$ magnitude for the HP\,1 RR Lyrae stars of $-0.11\pm0.14$. 
Random and systematic uncertainties in these calibrations were taken into account. 
A metallicity of [Fe/H]$=-1.06\pm0.10$ and a mean period of $0.42\pm0.04$ days were considered to compute this value. Therefore, the expected apparent distance modulus in $K_{\rm S}$ 
of HP\,1 is $14.47\pm0.18$, as shown in Figure \ref{RRLyr_Ks} (bottom panel). Subtracting from this value the expected $A_{K_{\rm S}}$ ($0.1184\times A_{V}=0.44\pm0.04$, assuming once again $E(B-V)=1.15\pm0.10$ and $R_{V}=3.2\pm0.10$), an intrinsic distance modulus of $(m-M)_{0}=14.03\pm0.18$ is recovered.
Converted to the heliocentric distance, this value means $6.4^{+0.6}_{-0.5}$ kpc.
It is interesting to note that this result is independent of the uncertainty in the $R_{V}$ value, and is almost unaffected by those in reddening.

Although the two distance determinations presented in this section agree within the errors, a shorter distance scale is clearly favoured because the result obtained with the NIR photometry present uncertainties that are significantly smaller than those obtained from the $V$-filter analysis.
Furthermore, a value of $6.4^{+0.6}_{-0.5}$ kpc is in good agreement with the previous determinations from NIR \citep{Valenti+10, Ortolani+11} and optical \citep{Ortolani+97}.

\subsection{RR Lyrae stars and He content}

RR Lyrae stars are core helium burning stars, thus an increase in luminosity is expected for those that are He-enhanced objects. Therefore, they can provide a good constraint to the He abundance, which in turn can help one to derive more reliable ages.  

A simple comparison between BaSTI models for HB stars with $Y\sim 0.25$ and $Y=0.30$ indicates that a helium enhancement of $\Delta Y\sim0.05$ (for [Fe/H]$\sim$-1.0) produces RR Lyrae stars that are $\sim$0.20 mag brighter in $V$. It means an increase of the same amount in terms of $(m-M)_{V}$ or $(m-M)_{0}$, toward more discrepant values, producing an unreliable heliocentric distance of 8.4 kpc for HP\,1. 

Very recently \citet{Lagioia+18} and Milone et al. (2018) found that the differences in He content between first and second stellar generations are small, amounting to $\Delta Y\simless$0.03.
As demonstrated by \citet{Nardiello+15}, such He-enhancement, if present, would be related to age differences of $\lesssim$ 300 Myr, even taking into account [Fe/H] and [$\alpha$/Fe] variations of 0.02 dex between the stellar generations.   

\section{Isochrone fitting}
\label{isot_fit}

\subsection{Our method}

To derive ages, metallicity, intrinsic distance moduli [$(m-M)_{0}$], and reddening values [$E(B-V)$] in an objective and self-consistent way, we performed isochrone fits using a Bayesian approach. 
For simplicity reasons, we assume that HP\,1 can be considered as a single stellar population (SSP), which is a very reliable hypothesis given its relatively low mass. Besides there is
no information in this respect, since this clusters has not been observed in ultraviolet filters 
cf. Piotto et al. (2015).  
Similar methods have been successfully tested and applied to analyse CMDs of open clusters \citep{Naylor+Jeffries06, Monteiro+10, Alves+12}, Galactic GC \citep{Hernandez+Valls-Gabaud08, Wagner-Kaiser+17, Kerber+18}, and stellar clusters in the Magellanic Clouds \citep{Dias+16b,Pieres+16,Perren+17}.
In brief, by computing the likelihood statistics, our method determines the isochrones that best reproduce the observed CMD. First, for the $j$th isochrone displaced by a given intrinsic distance modulus and reddening vector, we compute the minimum distance of the $i$th observed star to it, defined as 

\begin{equation}
$$ r_{ij} = {\rm{min}} \left[\left(\frac{{\rm colour}_{i} - {\rm colour}_{j}}{\sigma_{{\rm colour},i}}\right)^{2} + \left(\frac{{\rm mag}_{i} - {\rm mag}_{j}}{\sigma_{{\rm mag},i}}\right)^{2} \right]^{1/2}
\end{equation}

\noindent
where $\sigma_{{\rm colour},i}$ and $\sigma_{{\rm mag},i}$ are the uncertainties in colour and magnitude of this observed star. Then, the probability of the $i$th observed star to belong to the $j$th isochrone is given by

\begin{equation}
p_{ij} \propto \frac{1}{\sigma_{{\rm colour},i}\,\sigma_{{\rm mag},i}}\exp\left( \frac{-r_{ij}^{2}}{2} \right) 
\end{equation}

\noindent
Finally, the logarithmic likelihood that the total observed stars ($N_{\rm obs}$) are draw from the $j$th isochrone can be written by

\begin{equation}
\ln \mathcal{L} = \ln \sum_{i=1}^{N_{\rm obs}} p_{ij} \propto -2 \sum_{i=1}^{N_{\rm obs}} r_{ij} - \sum_{i=1}^{N_{\rm obs}} \ln (\sigma_{{\rm colour},j}) - \sum_{i=1}^{N_{\rm obs}} \ln (\sigma_{{\rm mag},j})
\end{equation}

\noindent
To test all acceptable solutions, we explore a wide and regular grid in the all parameter space, where the center corresponds to the values used in our control experiments (see Section \ref{control}). 
The grid we employed typically covers $\Delta$Age$=5.0$ Gyr, $\Delta E(B-V)=0.360$, $\Delta (m-M)_{0}=0.90$ and $\Delta$[Fe/H]$=0.40$ dex, in steps of 0.2 Gyr, 0.008 mag, 0.02 mag and 0.02 dex, respectively.
We applied the Markov chain Monte Carlo (MCMC) sampling technique to obtain the final parameters, as well as to study the confidence intervals and correlations between them. For this purpose, we used the \texttt{emcee} code \citep{Foreman-Mackey+13} to sample the posterior probability in the four-dimensional parameter space, assuming a uniform prior probability distribution within the acceptable physical ranges. 

The final solutions correspond to the sum of the posterior probability distributions over 10 MCMC samples, in order to minimize the stochastic effects of the Monte Carlo methods. The uncertainties in each parameter depend on several factors, such as the photometric errors, the photometric depth, the contamination by field stars, the filters choice, and even the quality of the stellar evolutionary models.
As can be seen by the sanity checks discussed in the next Section and the results presented in Sect. \ref{results}, in this work the typical formal uncertainties from the MCMC for age, [Fe/H], $(m-M)_{0}$ and $E(B-V)$ are $\sim$ 0.6 Gyr, 0.07 dex, 0.04 mag and 0.03 mag, respectively.
To improve the quality of the fit we restrict our analysis to the stars within 3.0 $\sigma$ from the median colour position for each magnitude bin, therefore minimizing the influence of the remaining field stars and other outliers that are not predicted by the isochrones (e.g., blue stragglers, binaries, etc).
All the details of the Python code developed by us to make the isochrone fitting will be presented in a future paper (Souza et al. in preparation).

\subsection{Control experiments}\label{control}

\begin{figure*}
\includegraphics[scale=0.50,angle=0]{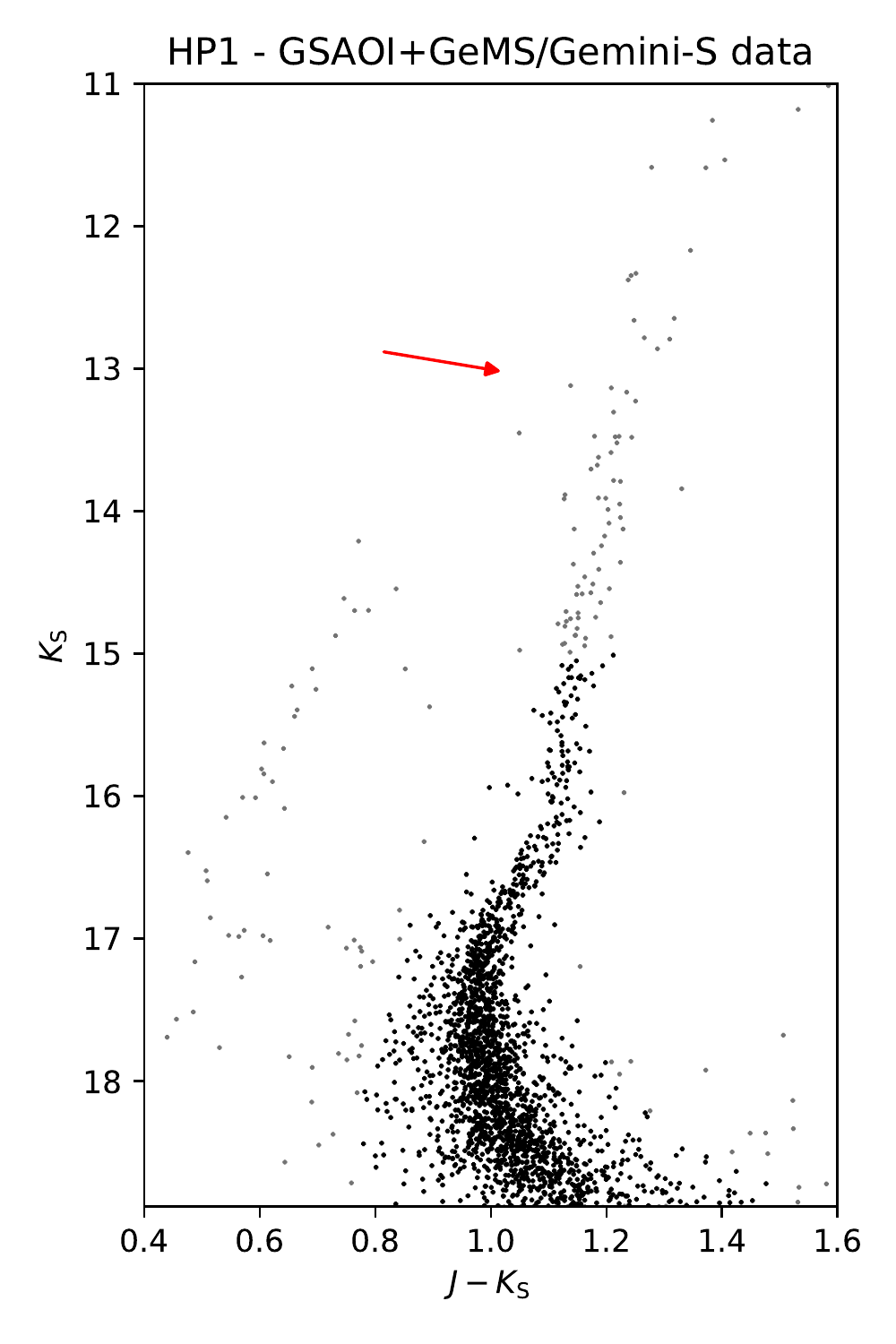}
\includegraphics[scale=0.50,angle=0]{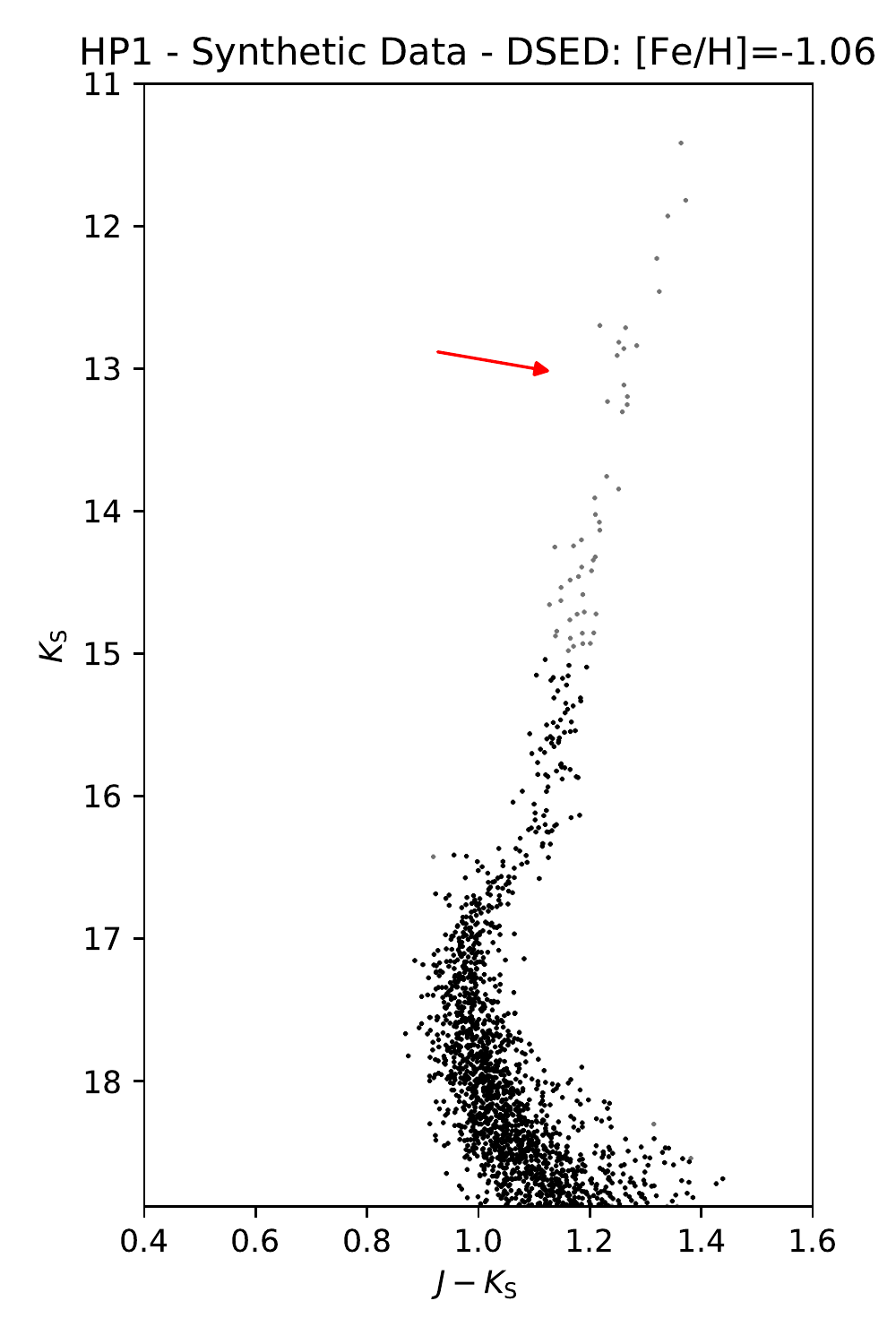}
\includegraphics[scale=0.40,angle=0]{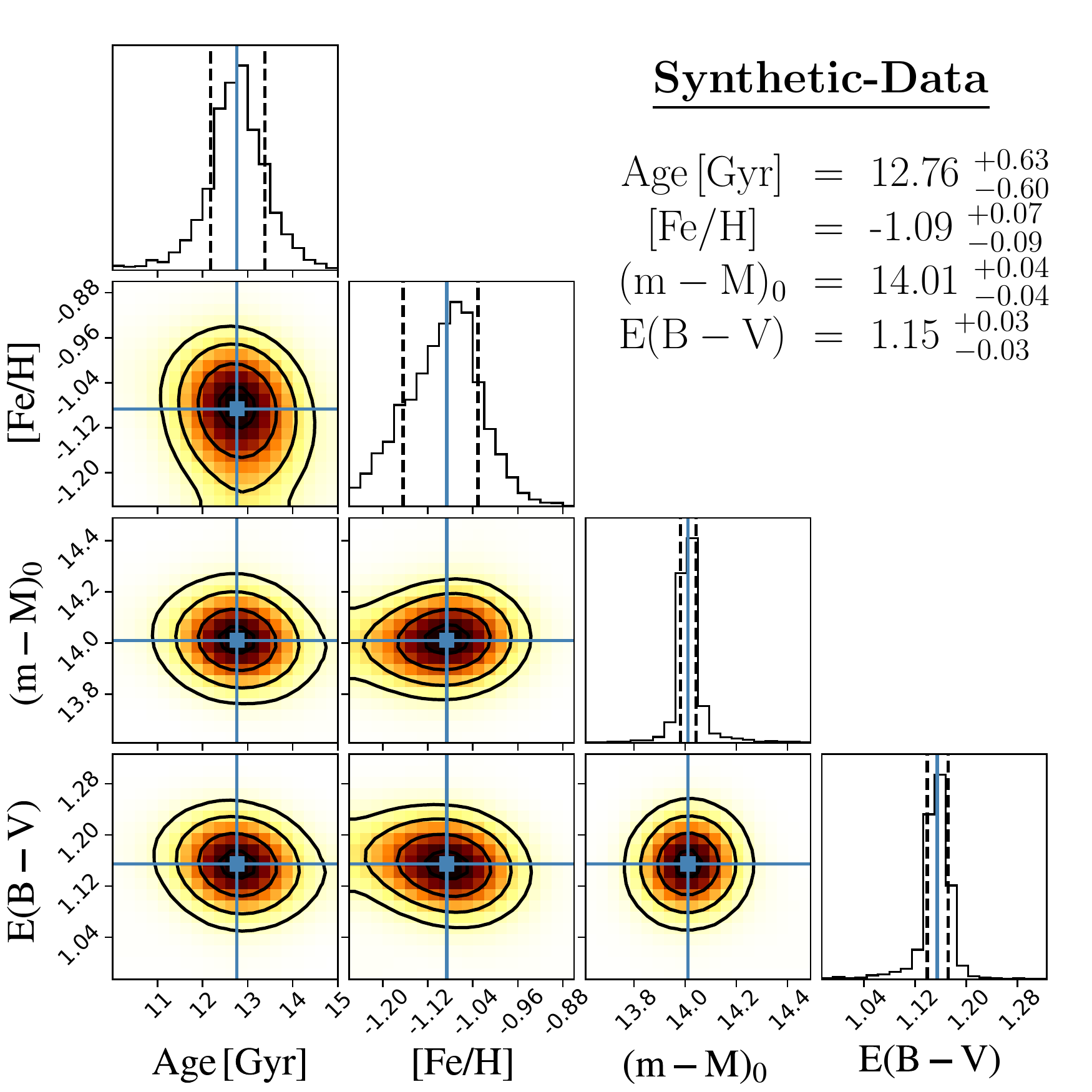}
    \caption{Results of a sanity check for our isochrone fitting technique in the $K_{\rm S}$ vs. $(J-K_{\rm S})$ CMD. Left panel: HP\,1 data. Middle panel: synthetic CMD of HP\,1 generated with DSED models. The input parameters are 12.5 Gyr, [Fe/H]$=-1.06$, $E(B-V)=1.15$, and $(m-M)_{0}=14.05$. 
    Only stars within 3.0$\sigma$ from the median colour position (black points) are used in the fit. Right panels: corner plots showing the output of the MCMC method and the recovered physical parameters. These panels present the one- and two-dimensional projections of the posterior probability distribution for all parameters. The contours correspond to the [0.5$\sigma$, 1.0$\sigma$, 1.5$\sigma$, 2.0$\sigma$] levels.}
    \label{sanity_test}
\end{figure*}

To validate our isochrone fitting technique, we performed some control experiments with synthetic CMDs. 
Artificial photometric catalogs mimicking the HP\,1 data were generated, providing synthetic data with known input parameters. 
This was done using a DSED isochrone with 12.5 Gyr, [Fe/H]$=-1.06$, $E(B-V)=1.15$, and $(m-M)_{0}=14.05$. 
A number of artificial stars similar to HP\,1 were randomly drawn assuming the observed photometric uncertainties and educated guesses for the binary fraction (30\%) and for the Initial Mass Function (Salpeter).
The results for a simple sanity check in the $K_{\rm S}$ vs. $(J-K_{\rm S})$ CMD are presented in Figure \ref{sanity_test},
where it is possible to attest that our method is recovering the expected parameters within the uncertainties. 
Similar successful results are also obtained for sanity checks in the optical-NIR CMD, as well as using BaSTI isochrones.

\begin{figure*}
\includegraphics[scale=0.30,angle=0]{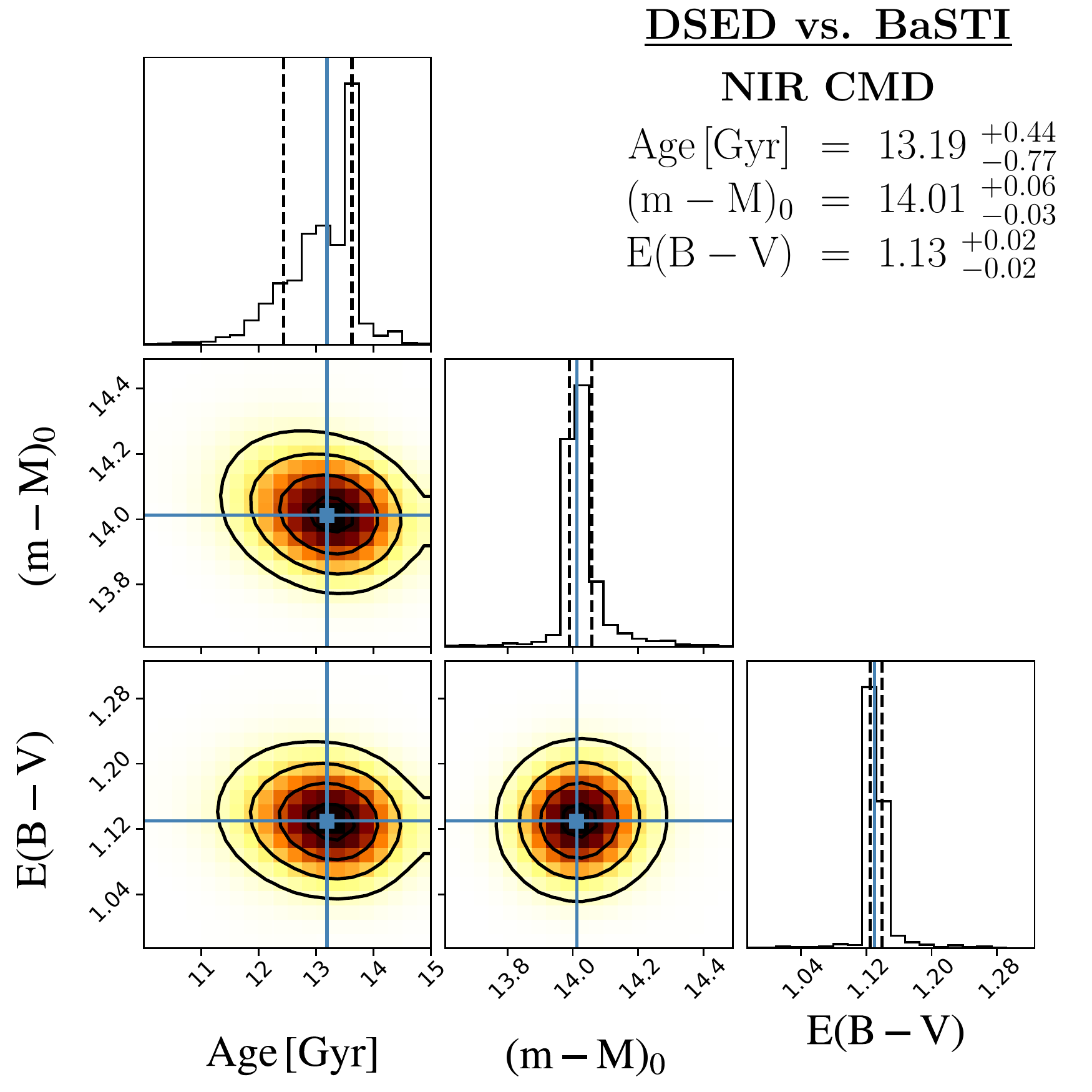}
\includegraphics[scale=0.30,angle=0]{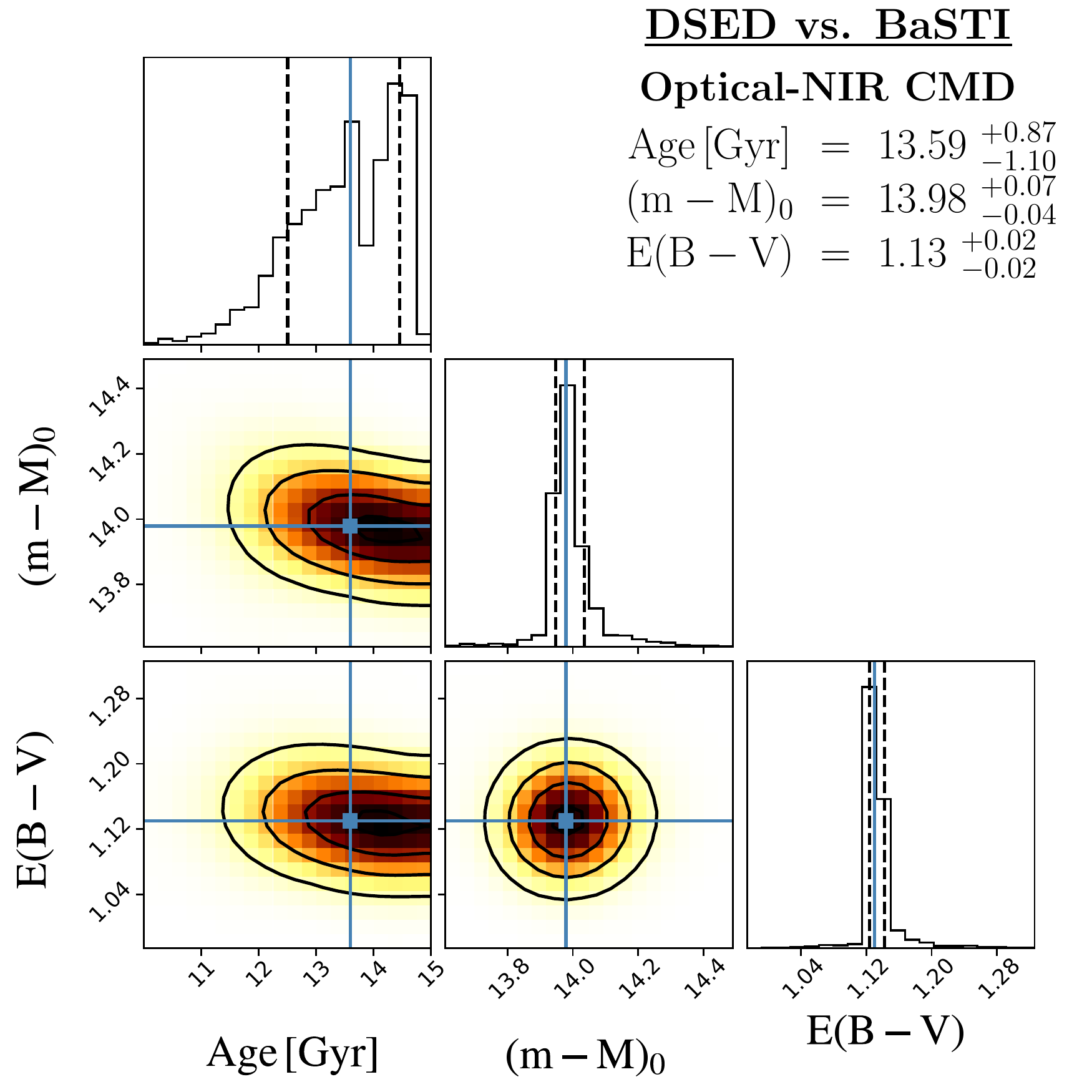}
\includegraphics[scale=0.38,angle=0]{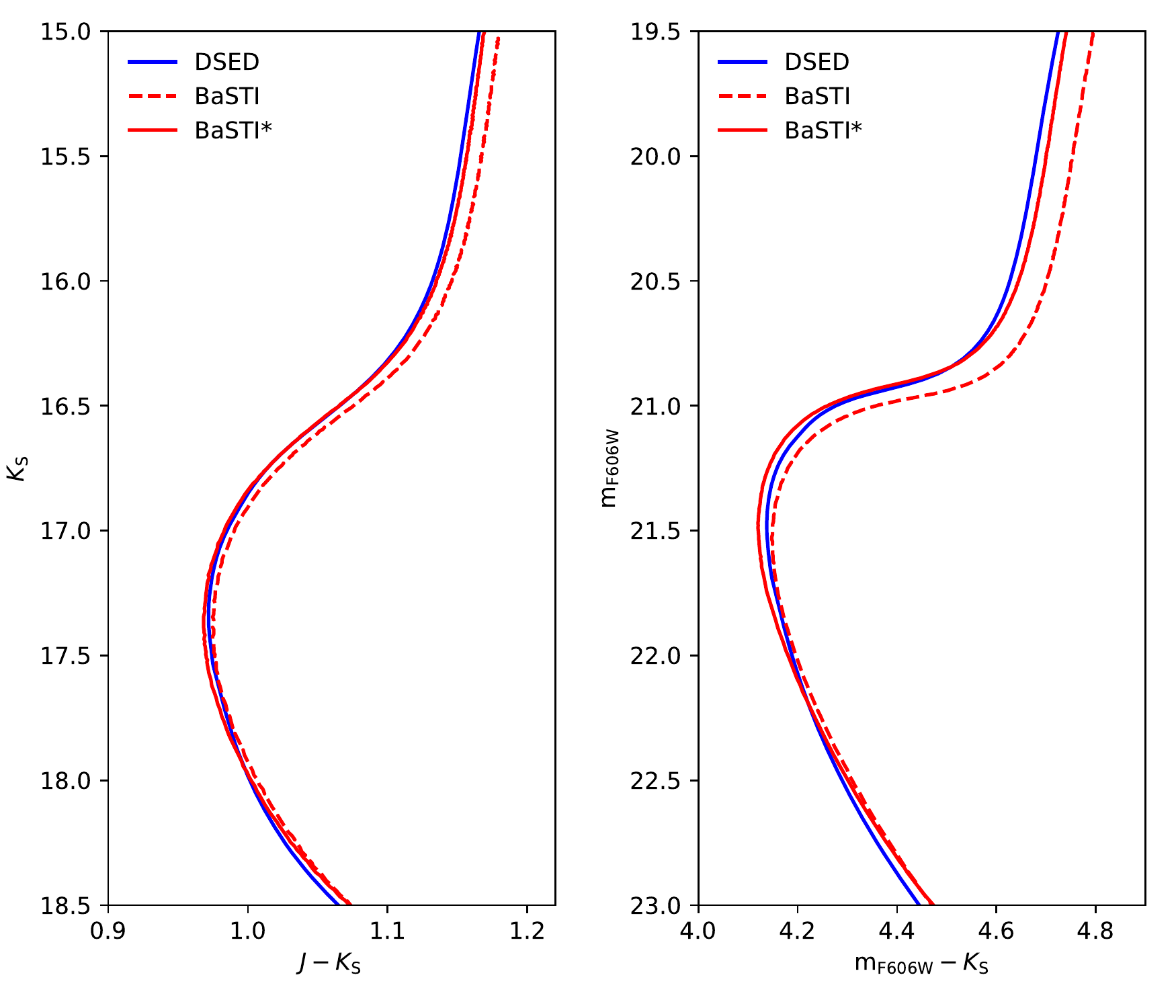}
    \caption{Results for two control experiments to quantify the expected differences in the physical parameters according with the stellar evolutionary choice. An DSED isochrone with 12.5 Gyr, [Fe/H]$=-1.06$, $E(B-V)=1.15$, and $(m-M)_{0}=14.05$ was used as input and compared with a regular grid of BaSTI isochrones. Left panels: Corner plots showing the output of the MCMC method for the $K_{\rm S}$ vs. $(J-K_{\rm S})$ CMD. Middle panels: results for the $m_{\rm F606W}$ vs. $(m_{\rm F606W}-K_{\rm S})$ CMD. They present the one- and two-dimensional projections of the posterior probability distribution for all parameters. The contours correspond to the [0.5$\sigma$, 1.0$\sigma$, 1.5$\sigma$, 2.0$\sigma$] levels. Right panels: the DSED isochrone used as input (blue solid line) and a BaSTI isochrone with the same parameters (red dashed line). We also present a BaSTI isochrone 1.0 Gyr younger and displaced by $\Delta E(B-V)=0.02$ and $\Delta (m-M)_{0}=0.05$ (BaSTI*).  }
    \label{dsed_vs_basti}
\end{figure*}

As discussed in Section \ref{isochrones}, it is expected that ages recovered using BaSTI isochrones will be older than the ones from fits with DSED isochrones. 
To quantify this effect and possible trends in the other parameters according with the choice in the stellar evolutionary models, we computed two additional control experiments.
Once again, a DSED isochrone with 12.5 Gyr, [Fe/H]$=-1.06$, $E(B-V)=1.15$, and $(m-M)_{0}=14.05$ was used as the input.
It is a convenient choice since the treatment for atomic diffusion is only present in the DSED code.
To avoid unnecessary stochastic effects, only the original isochrone magnitudes and colours were employed. 
By comparing this DSED isochrone with a regular and wide grid in the parameters space using BaSTI isochrones, we were able to determine the expected differences in the physical parameters. Note that in these tests we forced the metallicity to be fixed at $-1.06$. 
Figure \ref{dsed_vs_basti} reveals that our isochrone fits using BaSTI models recover solutions that are on average $\sim 0.9$ Gyr older, $\sim 0.02$ less red, and $\sim 0.05$ closer (in intrinsic distance modulus) than the ones with DSED models. 
In Section \ref{results}, we will apply the opposite of these offsets in order to correct the BaSTI solutions for atomic diffusion (hereafter, BaSTI*), therefore generating final values for the physical parameters that can be better compared. 

\section{Results from isochrone fitting}
\label{results}

\subsection{The $K_{\rm S}$ vs. $(J-K_{\rm S})$ CMD}

Figures {\ref{DSED_bestfit_k_jk} and \ref{BaSTI_bestfit_k_jk} present the best isochrone fits in the $K_{\rm{s}}$ vs. $J-$K$_{\rm{s}}$ CMD, using DSED and BaSTI models. 
In each of these figures we present an overview of the best solution with zoom showing the stars that were effectively used during the fit (left panel) and the output of the MCMC method (right panels). 
The best fits provide excellent solutions for both stellar evolutionary models, as attested by the overplotted isochrone and by the well defined posterior probability distributions in the corner plots. 
The recovered physical parameters for all fits using the NIR CMD are presented in the upper part of Table \ref{tab_all_parameters_GSAOI_ACS+GSAOI}. 
According to the analysis using DSED models, the age of HP\,1 from this CMD is $13.0^{+0.9}_{-0.6}$ Gyr, with an intrinsic distance modulus of $14.06 \pm 0.05$, which means a distance of $6.49^{+0.15}_{-0.15}$ kpc, a reddening value of $E(B-V)=1.13\pm0.03$ and a metallicity of [Fe/H]$= -1.07^{+0.07}_{-0.10}$.
It is important to highlight that only assume a uniform prior probability distribution for metallicity during the isochrone-fitting process, and even so the recovered value is in very good agreement with the one estimated with high-resolution spectroscopy \citep{Barbuy+16}.

A very interesting consistent check can be done by comparing the apparent distance modulus from the isochrone fits with those from the RR Lyrae stars. 
The DSED models predict a value of $14.49\pm 0.05$ for the apparent distance modulus in $K_{\rm S}$, in perfect agreement with the one determined in Section \ref{RRLyrae} (See Figure \ref{RRLyr_Ks}). 
It is important to note that these two results are totally independent and based on pure NIR data, reinforcing a short scale distance for HP\,1. 

The original results with BaSTI models point to a very old age of $13.8^{+0.9}_{-1.2}$ Gyr, being in the limit of the age of the Universe \citep[13.799$\pm0.021$ Gyr,][]{Planck16}.
This age is reduced to $12.9^{+0.9}_{-1.2}$ Gyr after the expected correction of 0.9 Gyr due to the atomic diffusion, putting the new value in very good agreement with the one from DSED.
Furthermore, the expected offsets for the other parameters leave the distance and reddening from BaSTI and DSED in a perfect match.

\subsection{The $m_{\rm F606W}$ vs. $(m_{\rm F606W}-K_{\rm S})$ CMD}

The best isochrone fits using the $m_{\rm{F606W}}$ vs. $(m_{\rm{F606W}}-K_{\rm{s}})$ CMD and DSED and BaSTI models 
are presented in Figures {\ref{DSED_bestfit_v_vk} and \ref{BaSTI_bestfit_v_vk}. The recovered physical parameters for all fits using the optical-NIR CMD are presented in the bottom part of Table \ref{tab_all_parameters_GSAOI_ACS+GSAOI}.
As in the pure NIR data, the best solutions reproduce the CMD features very well. 
The corner plots from the MCMC method present well defined one- and two-dimensional projections of the posterior probability distributions for all parameters \citep{Foreman-Mackey+16}, attesting the high quality of the fits.

The results from DSED isochrones in this optical-NIR CMD indicate an age of 12.71$^{+0.7}_{-0.54}$ Gyr, in excellent agreement with the one presented for in the previous subsection. 
Although a slightly higher intrinsic distance modulus (or distance) is recovered here, this value agrees within 1$\sigma$ with the one from the analysis of GSAOI+GeMS data. 
The same cannot be said for the reddening: it is clearly higher when the optical \textit{HST} data is introduced in the analysis. With respect to the metallicity, the recovered value is in good agreement both in the fitting with the NIR data and with the spectroscopic analysis.  

Once again the RR Lyrae provide an independent check concerning the apparent distance modulus. 
By summing the $(m-M)_{0}$ with $3.2\times E(B-V)$, we estimated $(m-M)_{V}$ as $17.80^{+0.08}_{-0.09}$. 
This result is in agreement 
with the one presented in Section \ref{RRLyrae} within the limit of $1\sigma$ confidence interval (See Figure \ref{RRLyr_V}). 

BaSTI isochrones essentially confirm the distance and reddening values from the analysis using DSED models.
However, the age from BaSTI after the expected correction is somewhat younger than all previous determinations, including those from the same stellar evolutionary models but analysing pure NIR data.  

\begin{figure*}
\includegraphics[width=0.9\columnwidth, scale=0.5]{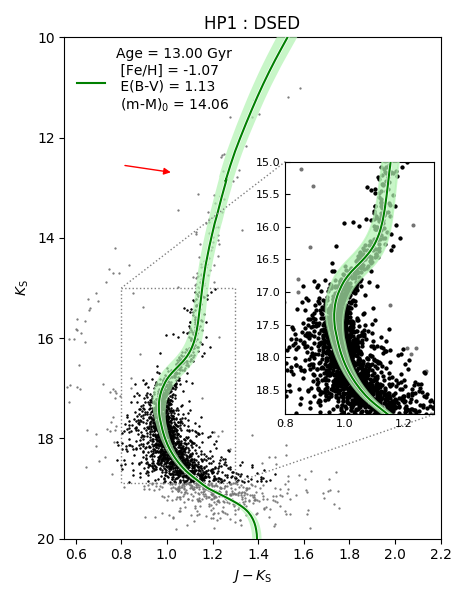}\;\;
\includegraphics[scale=0.51,angle=0]{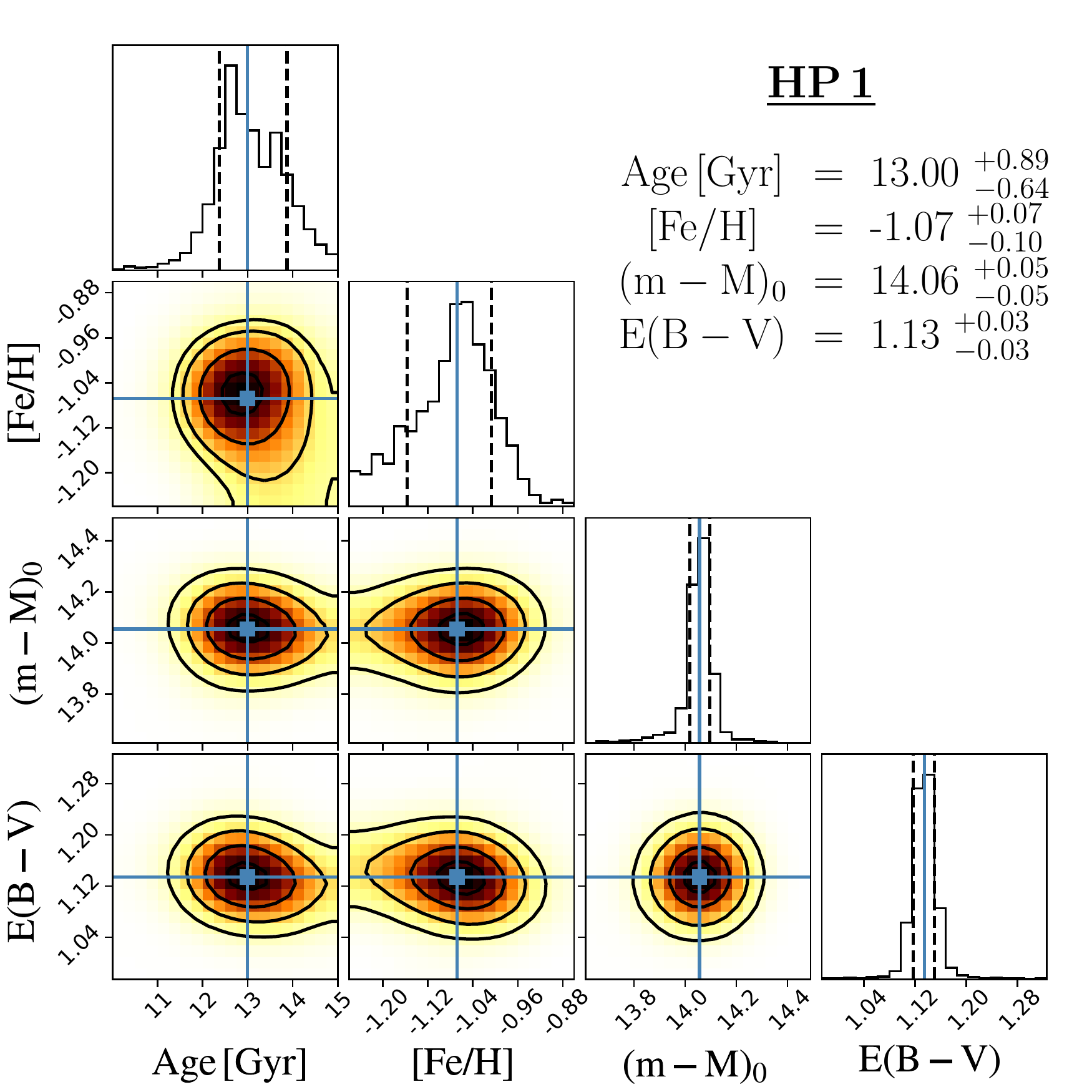}
\caption{Best isochrone fit in the $K_{\rm{s}}$ vs. $(J-K_{\rm{s}})$ CMD using DSED models. Left panel: CMD showing all observed stars (gray) and those used in the fit (black). A reddening vector corresponding to $\delta E(B-V)=0.10$ is also presented, and a zoom in the CMD region that is more sensitive to the age. The best fit is highlighted by a thick line. The green area shows the region between the two isochrones using the values within 1.0$\sigma$ for all the parameters. Right panels: corner plots showing the output of the MCMC method. They present the one- and two-dimensional projections of the posterior probability distribution for all parameters. The contours correspond to the [0.5$\sigma$, 1.0$\sigma$, 1.5$\sigma$, 2.0$\sigma$] levels.
 }
    \label{DSED_bestfit_k_jk}
\end{figure*}

\begin{figure*}
\includegraphics[width=0.9\columnwidth, scale=0.5]{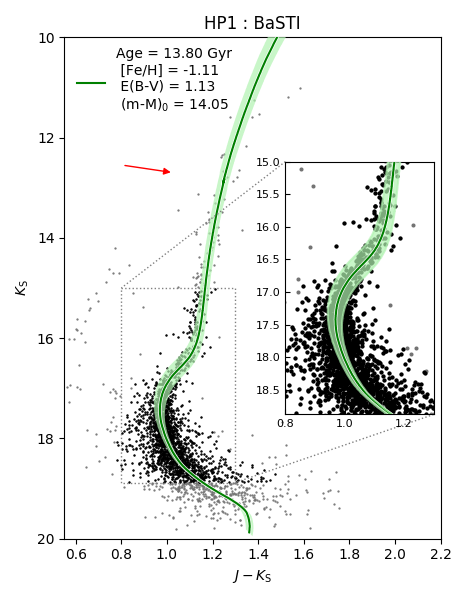}\;\;
\includegraphics[scale=0.51,angle=0]{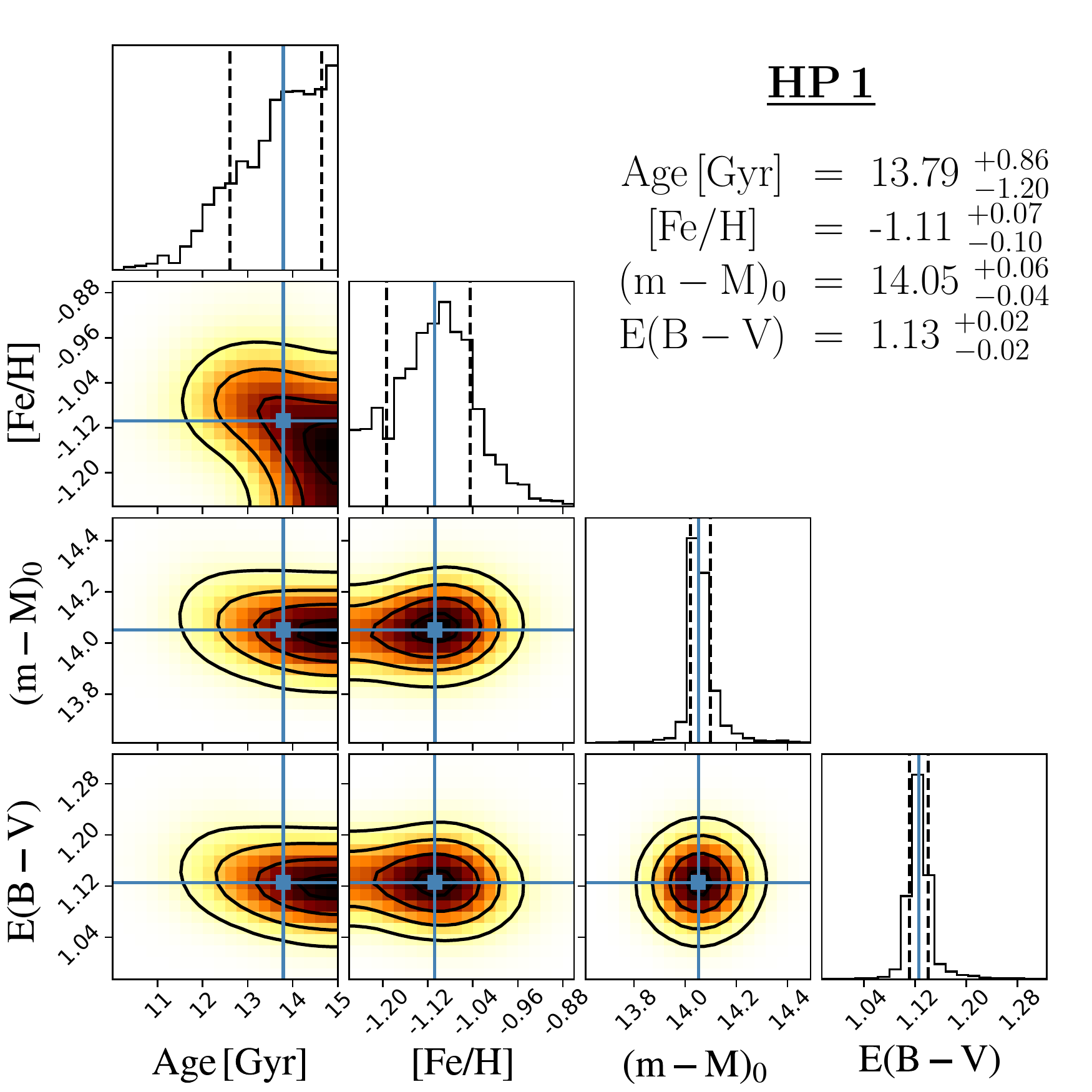}
   \caption{Same as Figure \ref{DSED_bestfit_k_jk}, but for BaSTI models.
 }
    \label{BaSTI_bestfit_k_jk}
\end{figure*}

\begin{figure*}
\includegraphics[width=0.9\columnwidth, scale=0.5]{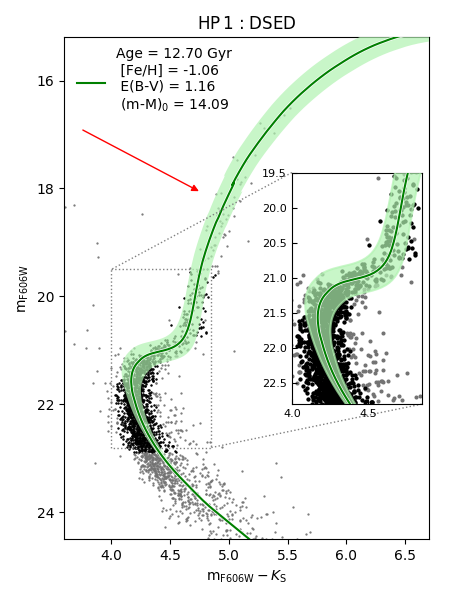}\;\;
\includegraphics[scale=0.51,angle=0]{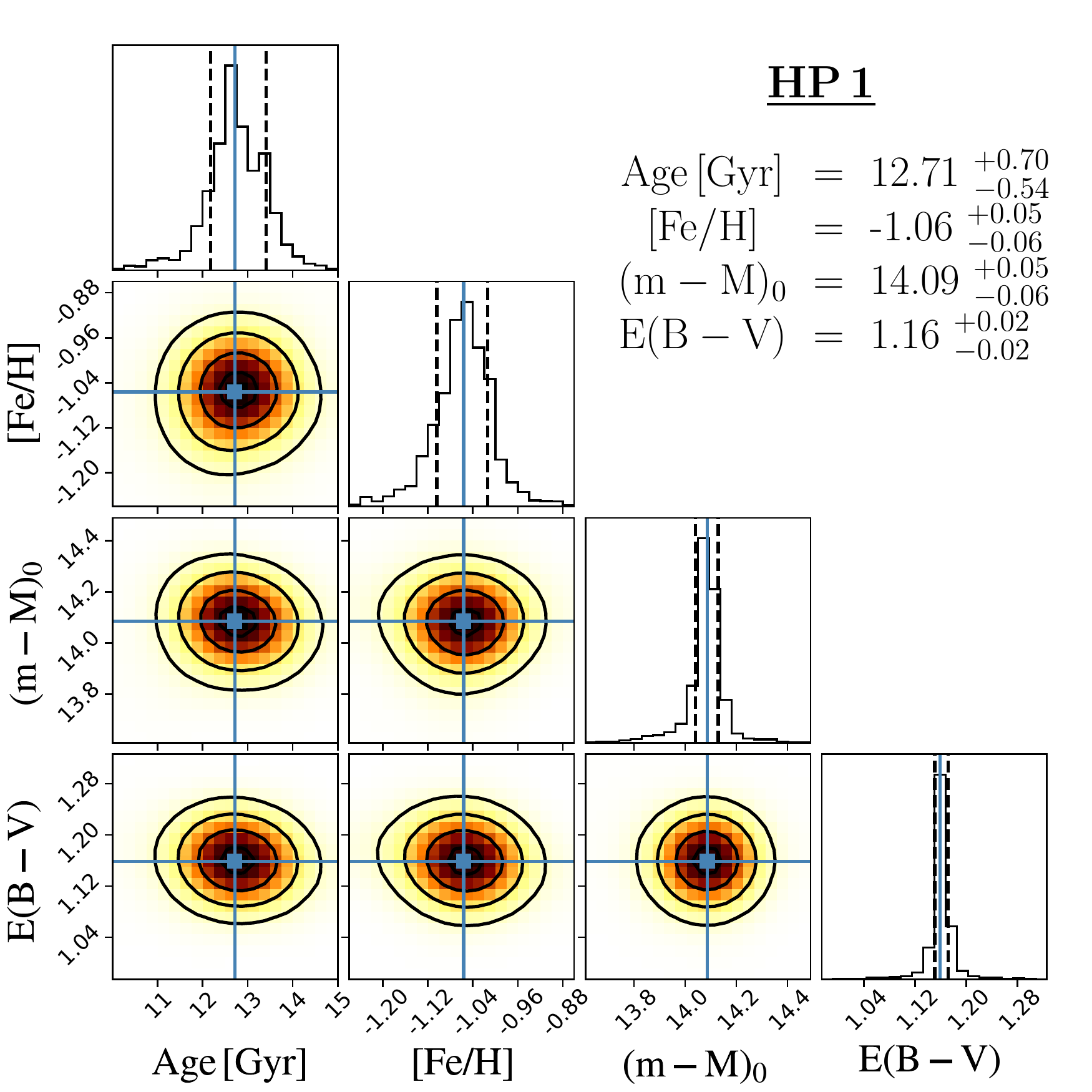}
	\caption{Same as Figure \ref{DSED_bestfit_k_jk} but for $m_{\rm{F606W}}$ vs. $(m_{\rm F606W}-K_{\rm{s}})$ CMD using DSED models and a reddening vector corresponding to $\delta E(B-V)=0.30$.
 }
    \label{DSED_bestfit_v_vk}
\end{figure*}

\begin{figure*}
\includegraphics[width=0.9\columnwidth, scale=0.5]{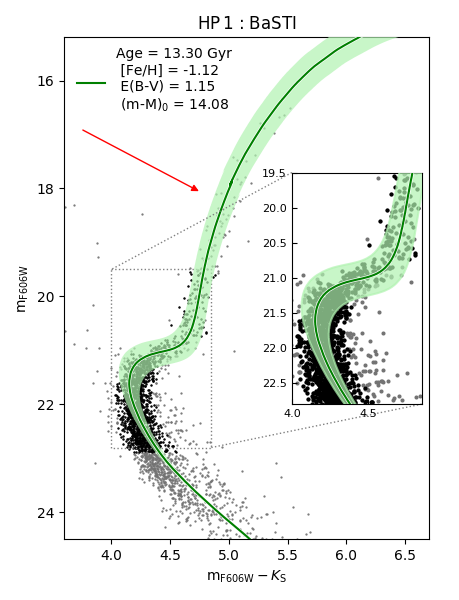}\;\;
\includegraphics[scale=0.51,angle=0]{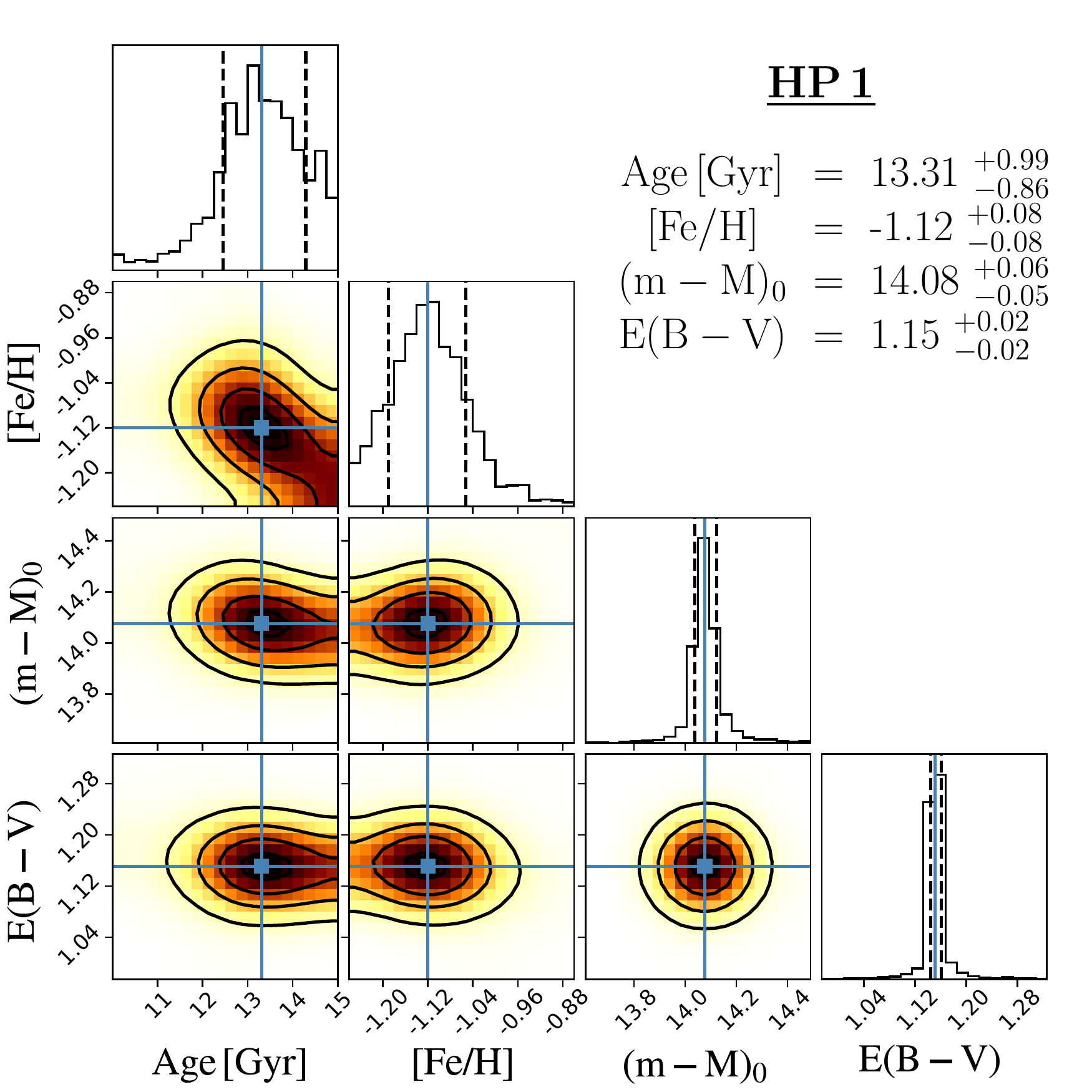}
    \caption{Same as Figure \ref{DSED_bestfit_v_vk}, but for BaSTI models.
 }
    \label{BaSTI_bestfit_v_vk}
\end{figure*}


\begin{table*}
\begin{center}
\caption{Physical parameters for HP\,1 from isochrone fits using $\alpha$-enhanced ([$\alpha$/Fe]=+0.40) BaSTI and DSED models in the $K_{\rm S}$ vs. $(J-K_{\rm S})$ and $m_{\rm F606W}$ vs. $(m_{\rm F606W}-K_{\rm S})$ CMDs.}
\begin{tabular}{lcccccc}
\hline
\hline
Model& [Fe/H] & Age   & $(m-M)_{0}$  & $d_{\odot}$  & $E(B-V)$ & $(m-M)^{\dag}$ \\ 
 & & (Gyr) & & (kpc) & &         \\ 
\hline
\multicolumn{7}{c}{$K_{\rm S}$ vs. $(J-K_{\rm S})$ CMD}\\
\hline
DSED  & $-1.07^{+0.07}_{-0.10}$ & 13.00$^{+0.89}_{-0.64}$ & 14.06$^{+0.05}_{-0.05}$ & 6.49$^{+0.15}_{-0.15}$ & 1.13$^{+0.03}_{-0.03}$ & 14.49$^{+0.05}_{-0.05}$ \\ [1ex]
BaSTI   & $-1.11^{+0.07}_{-0.10}$ & 13.79$^{+0.86}_{-1.20}$ & 14.05$^{+0.06}_{-0.04}$ & 6.46$^{+0.18}_{-0.12}$ & 1.13$^{+0.02}_{-0.02}$ & 14.48$^{+0.0}_{-0.0}$ \\ [1ex]
BaSTI$^{*}$ & $-1.11^{+0.07}_{-0.10}$ & $12.89^{+0.86}_{-1.20}$ & $14.10^{+0.06}_{-0.04}$ & $6.61^{+0.19}_{-0.12}$ & $1.15^{+0.02}_{-0.02}$ & $14.54^{+0.06}_{-0.04}$ \\ [1ex]
\hline
\multicolumn{7}{c}{$m_{\rm F606W}$ vs. $(m_{\rm F606W}-K_{\rm S})$ CMD}\\
\hline
DSED  & $-1.06^{+0.05}_{-0.06}$ & 12.71$^{+0.70}_{-0.54}$ & 14.09$^{+0.05}_{-0.06}$ & 6.58$^{+0.15}_{-0.18}$ & 1.16$^{+0.02}_{-0.02}$ & 17.80$^{+0.08}_{-0.09}$ \\ [1ex]
BaSTI   & $-1.12^{+0.08}_{0.08}$ & 13.31$^{+0.99}_{-0.86}$ & 14.08$^{+0.06}_{-0.05}$ & 6.55$^{+0.18}_{-0.15}$ & 1.15$^{+0.02}_{-0.02}$ & 17.76$^{+0.09}_{-0.08}$ \\ [1ex]
BaSTI$^{*}$ & $-1.12^{+0.08}_{+0.08}$ & $12.41^{+0.99}_{-0.86}$ & $14.13^{+0.06}_{-0.05}$ & $6.70^{+0.19}_{-0.16}$ & $1.17^{+0.02}_{-0.02}$ & $17.87^{+0.09}_{-0.08}$ \\ [1ex]
\hline
\multicolumn{7}{c}{Average results (DSED \& BaSTI* models)}\\
\hline
 &$-1.09^{+0.07}_{-0.09}$&$12.75^{+0.86}_{-0.81} $& $14.10^{+0.06}_{-0.05}$ & $6.59^{+0.17}_{-0.15}$& $1.15^{+0.02}_{-0.02}$& $-$ \\
\hline
\hline
\end{tabular}
\label{tab_all_parameters_GSAOI_ACS+GSAOI}
\\
\end{center}
BaSTI$^{*}$ is the BaSTI results after the expected correction for atomic diffusion (See Sections \ref{isochrones} and \ref{isot_fit} for details).
\\$^{\dag}$$(m-M)$ is the apparent distance modulus in the $K_{\rm{s}}$ filter (for the NIR CMD)  or in the $V$ filter (for the optical-NIR CMD).
\end{table*}


\section{Discussion on Age and Distance}
\label{discussion}

From the previous section, it is clear that there is an excellent agreement between the ages and distances for both stellar evolutionary models and both CMDs. 
Taking into account the atomic diffusion, the average values for the age and distance are $12.8^{+0.9}_{-0.8}$ Gyr and $6.59^{+0.17}_{-0.15}$ kpc. The uncertainties in these values are largely dominated by the formal ones from the MCMC method. The systematics uncertainties, i.e., those due to the different choices of model and CMD, are significantly smaller, as attested by the standard deviation of the mean over the four determinations (0.2 Gyr and 0.05 kpc). 

The absolute ages recovered in the present work locate HP\,1 as one of the most ancient GCs in our Galaxy, probably formed in the first Gyr of the Universe. 
This corroborates the hypothesis that relatively metal-poor bulge GCs ([Fe/H]$\lesssim-1.0$) with BHB and $\alpha$-enhancement can be older than 12.5 Gyr, as very recently demonstrated for NGC 6558 \citep{Barbuy+18}}, NGC\,6522 and NGC\,6626 \citep{Kerber+18} using optical proper-motion-cleaned CMDs obtained with \textit{HST}.
Although metal-poor halo GCs ([Fe/H]$\sim -2.0$), like M15 \citep[e.g.,][]{Monelli+15}, are the most obvious candidates to harbour the oldest stars in the MW, we are demonstrating that a very low metallicity is not a restrictive prior condition to find very old stars in the bulge. 
 In fact, the existence of some GCs with $-1.4 \lesssim$ [Fe/H] $\lesssim -1.0$ and $R_{\rm GC} \lesssim 3.0 $ kpc among the oldest GCs was reported by several works using $HST$ data \citep{Marin-Franch+09, Dotter+10, VandenBerg+13,Wagner-Kaiser+17}.  
Furthermore, \citet{Dotter+10} identify the age as the second parameter to explain the HB morphology, being the clusters with BHB the oldest ones.
A notable example of such old GC with all aforementioned characteristics is NGC\,6717.

In order to compare our derived age results for HP\,1, we here discuss two recent papers that determined absolute ages for GCs using deep NIR data collected with GSAOI+GeMS@Gemini-S. 
\citet{Saracino+16} presented the analysis of a $K_{\rm S}$ vs $(J-K_{\rm S})$ CMD of NGC\,6624, a bulge GC with [Fe/H]$=-0.69\pm0.02$ and [$\alpha$/Fe]$\sim+0.39$ \citep{Valenti+11}. 
Using three different stellar evolutionary models, including DSED and BaSTI, they recovered an age of $12.0\pm0.5$ Gyr, somewhat younger than the one determined by us for HP\,1. 
These authors did not apply the correction for the lack of atomic diffusion in the BaSTI models, therefore overestimating the age of NGC\,6624. 
A similar age of $11.9\pm0.7$ (intrinsic)$\pm0.45$ (metallicity term) Gyr was determined for NGC\,2808 by \citet{Massari+16}, a halo GC with [Fe/H]$=-1.13\pm0.04$ \citep{Carretta15}, i.e., a GC with a metallicity similar to the one of HP\,1. 
Combining the NIR data from GSAOI+GeMS with optical photometry extracted from \textit{HST} images, they estimated the ages computing the difference between the MSTO and the MS knee presented in the NIR.
Curiously, NGC\,6624 and NGC\,2808 present simultaneously blue and red HB stars, as well as remarkable multiple populations when UV and optical \textit{HST} images are employed to build CMDs \citep{Piotto+15}. 
Furthermore, NGC\,2808 presents a complex GC with five distinct stellar populations according to the [Na/Mg] ratios \citep{Carretta15, Milone+15}. Despite these features, the stars within each of these clusters have similar ages, probably not so old as the ones in HP\,1.

Since we have a very accurate heliocentric distance, directly confirmed from the analysis of RR Lyrae stars in the NIR, we can estimate a Galactocentric distance ($R_{\rm GC}$}) for HP\,1 only limited by the uncertainties in the distance of the Sun to the Galactic center ($R_{0}$).
Assuming $R_{0}=8.2\pm0.2$ kpc from the recent statistical analysis done by \citet{Bland-Hawthorn2016} over 26 independent determinations, we recover a value of $R_{\rm GC}=1.7\pm0.3$ kpc for HP\,1. 

\section{Orbital Analysis of HP\,1}
\label{orbit}

We determined a new, accurate (uncertainties $\lesssim 2$ \%) distance
of HP\,1. We used this distance, combined with the absolute PMs and the
radial velocity, of the cluster, to refine the orbit of HP\,1 computed in \citet{Ortolani+11}.

Our PMs are relative to the bulk motion of the cluster and not in an
absolute reference-frame system. Therefore, we made use of the
absolute PMs available in the Gaia DR2 catalog
\citep{Gaia16a,Gaia18} to compute the bulk
absolute motion of HP\,1.

First, we selected all stars within 0.1 deg
from the center of HP\,1 with an absolute PM measurement. Then, we rejected all
objects with a poorly-measured absolute PM according to the prescriptions given
in \citet{Arenou+18}. We further refined the sample by
considering only stars with an absolute PM error smaller than 0.25 mas
yr$^{-1}$. Finally, we computed the 5$\sigma$-clipped median value of
the absolute PMs along $\alpha \cos\delta$ and $\delta$ directions. We find:
\begin{equation*}
  \begin{array}{ll}
    (\mu_\alpha\cos\delta,\mu_\delta)_{\rm HP\,1} 
    = (2.409\pm0.042,-10.136\pm0.031)~{\rm mas~yr^{-1}} \, .\\
  \end{array}
\end{equation*}
The error bars do not include the systematic error of 0.035 mas
yr$^{-1}$ described in \citet{Gaia18}. Our PMs are consistent with the PMs determined by \citet{Vasiliev18} and \citet{Baumgardt+18}, both based on Gaia DR2 PMs as well. 

The registration of our relative PMs in an absolute system was not
possible because of the large magnitude difference between our and the Gaia catalogs. Therefore, we adopted the Gaia-based absolute PMs as input
parameter for the orbit computation.

The radial velocity, from high resolution spectroscopy, adopted for the computation is $40.0 \pm 0.5$ km s$^{-1}$ \citep{Barbuy+16}. Interestingly, we found one star with a radial-velocity measurement in the Gaia DR2 catalog that is member of HP1 from the PMs. The radial velocity of this probable HP1 star is $42.25 \pm 2.75$ km s$^{-1}$. This value is in agreement with that computed by \citet{Barbuy+16}.


The axisymmetric Galactic model includes a S\'ersic bulge, an exponential disc generated by the superposition of three Miyamoto-Nagai potentials \citep{Miyamoto+75}, following the methodology made by \citet{Smith+15}, and a dark matter halo is modelled with the Navarro-Frenk-White (NFW) density profile \citep{Navarro+97}, having a circular velocity $V_0=241$ km s$^{-1}$ at $R_0=8.2$ kpc \citep{Bland-Hawthorn2016}. For the Galactic bar, we employed a triaxial Ferrer's ellipsoid, where all the mass from the S\'ersic bulge component is converted into bar. For the bar potential, we assume a total bar mass of $1.2 \times 10^{10}$ M$_{\odot}$, an angle of $25^{\circ}$ with the Sun-major axis of the bar, a gradient of pattern speed of the bar $\Omega_b= 40$, 45, and 50 km s$^{-1}$ kpc$^{-1}$, and a major axis extension of 3.5 kpc. Even though we vary the bar angular velocity, we keep the same bar extension. In our Galactic model we are not taking into account the contribution of the spiral arms, given that HP\,1 is confined inside 3 kpc from the Galactic centre, therefore the spiral arms would have a negligible effect.

The integration of the orbits was made with the \texttt{NIGO} tool \citep{Rossi15a}, which includes the potentials mentioned above. The solution of the equations of motion is evaluated numerically using the Shampine--Gordon algorithm (for details, see \citealt{Rossi15b}). We adopted the right-handed, Galactocentric Cartesian system, $x$ toward the Galactic centre, and $z$ toward the Galactic North Pole. The initial conditions of the cluster is obtained from the observational data, coordinates, heliocentric distance, radial velocity, and absolute proper motions given in Table \ref{tab:HP1_para}. The velocity components of the Sun with respect to the local standard of rest are $(U,V,W)_{\odot}= (11.1, 12.24, 7.25)$ km s$^{-1}$ \citep{Schonrich+10}. In order to evaluate the effect of the uncertainties associated to the HP\,1's parameters, we employ the Monte Carlo method to generate a set of 1000 initial conditions taking into account the errors of distance, heliocentric radial velocity and absolute proper motion components. We integrate the orbits with such initial conditions forward for 10 Gyr. For each orbit, we calculate the perigalactic distance $r_{\rm min}$, apogalactic distance $r_{\rm max}$, the maximum vertical distance from the Galactic plane $|z|_{\rm max}$, and the eccentricity defined by $e=(r_{\rm max}-r_{\rm min})/(r_{\rm max}+r_{\rm min})$.

\begin{table}
\caption{HP\,1 parameters for the orbit integration.}
\begin{tabular}{@{}lcc@{}} 
\hline
Parameter & Value & Reference\\
\hline
$(\alpha,\delta)_{(J2000)}$ & ($17^{\rm h}31^{\rm m}05^{\rm s}.2$, $-29^{\circ}$58'54'' ) & 1 \\
Radial velocity & $40\pm 0.5$ km s$^{-1}$& 2 \\
d$_{\odot}$ & $6.59\pm 0.16$ kpc & This work  \\
$\mu_{\alpha} \cos \delta$ &$2.409\pm 0.054^{\dag}$ mas yr$^{-1}$& This work$^*$ \\
$\mu_{\delta}$ & $-10.136 \pm 0.046^{\dag}$  mas yr$^{-1}$& This work$^*$\\
\hline
\end{tabular}
\label{tab:HP1_para}
\\References --- (1) \citealt{harris96}. (2) \citealt{Barbuy+16}. 
\\$^{\dag}$Uncertainty includes the systematic error of 0.035 mas yr$^{-1}.$
\\$^*$ Obtained by using Gaia DR2 \citep{Gaia18}.
\end{table}

\begin{figure}
	\includegraphics[width=\columnwidth]{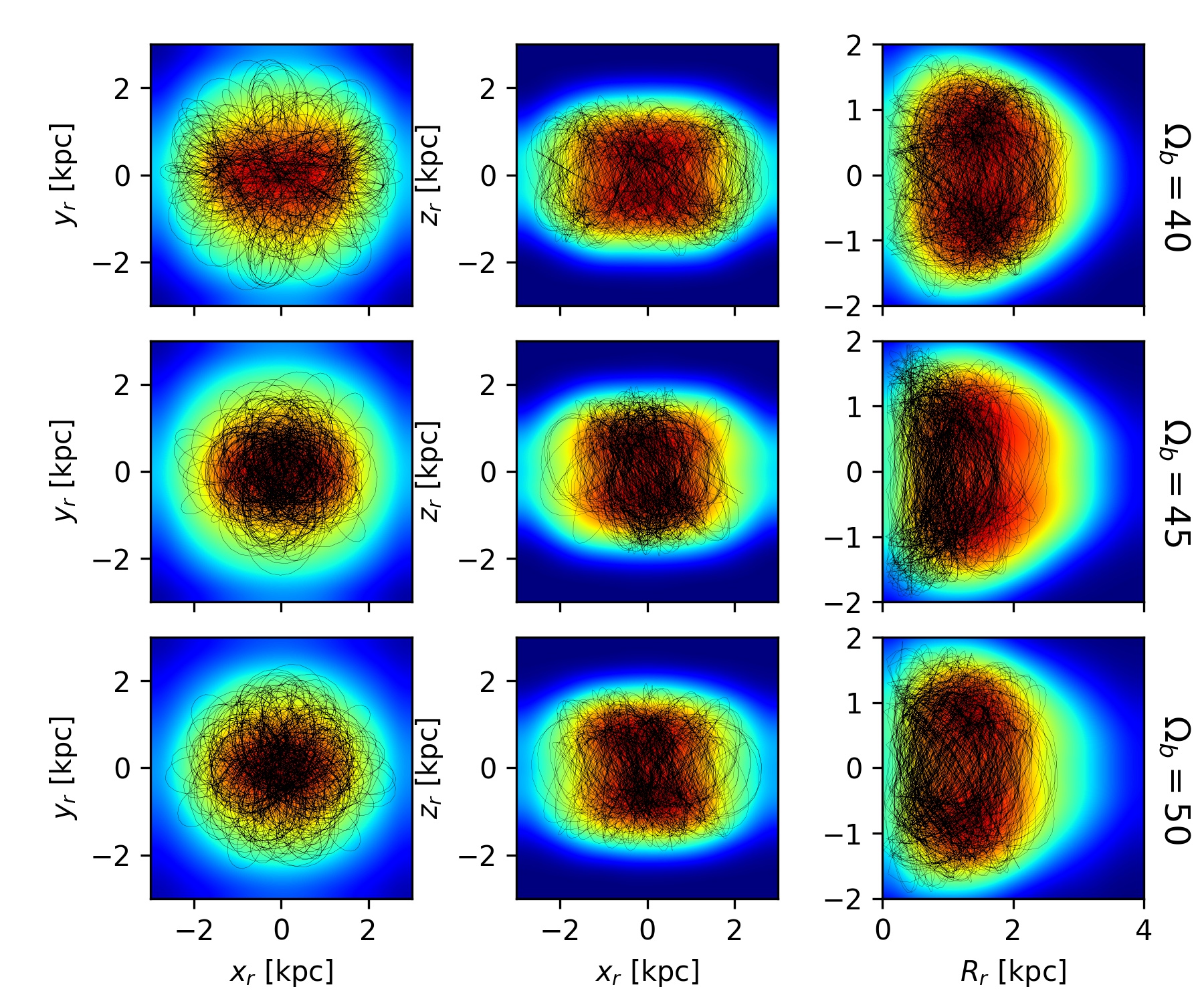}
    \caption{Probability density map for the $x_r-y_r$ (left panels), $x_r-z_r$ (middle panels), and $R_r-z_r$ (right panels) projections of the 1000 Orbits for HP\,1 co-rotating with the bar frame, with a gradient in the angular velocity of the Galactic bar: $40$ (top row), $45$ (middle row) and $50$ (bottom row) km s$^{-1}$ kpc$^{-1}$. The black lines show the orbits using the central values given in Table \ref{tab:HP1_para}.  
    }   
    \label{orbits}
\end{figure}

Figure \ref{orbits} shows the probability densities of the orbits of HP\,1, in the frame co-rotating with the bar. For each value of angular velocity we analysed, we show the $x_r-y_r$, $x_r-z_r$, and $R_r-z_r$ projections (different rows). The red and yellow colours highlight the space region that the orbits cross more frequently. The orbit with central values given in Table \ref{tab:HP1_para} (black curve) is confined in the inner region of the Galaxy, inside 3 kpc, having a boxy-shape orbit in $x_r-z_r$ projection with an almost negligible dependence of the angular velocity. This kind of orbits give the orbital support to the X-shape in the Galactic bulge \citep{Portail+15}. For the case of $\Omega_b = 40$ km s$^{-1}$ kpc$^{-1}$, the orbit reaches closer to the Galactic centre. Additionally, we calculate the z-angular momentum in the inertial reference frame, $Lz$, with the purpose of checking its sign, given that $Lz$ is not conserved, to determine the sense of motion of HP 1 orbit. We found that the orbit of HP\,1 has prograde and retrograde motion at the same time. This has been connected with chaotic behaviour \citep{Pichardo+04}.     

\begin{figure*}
	\includegraphics[width=17cm]{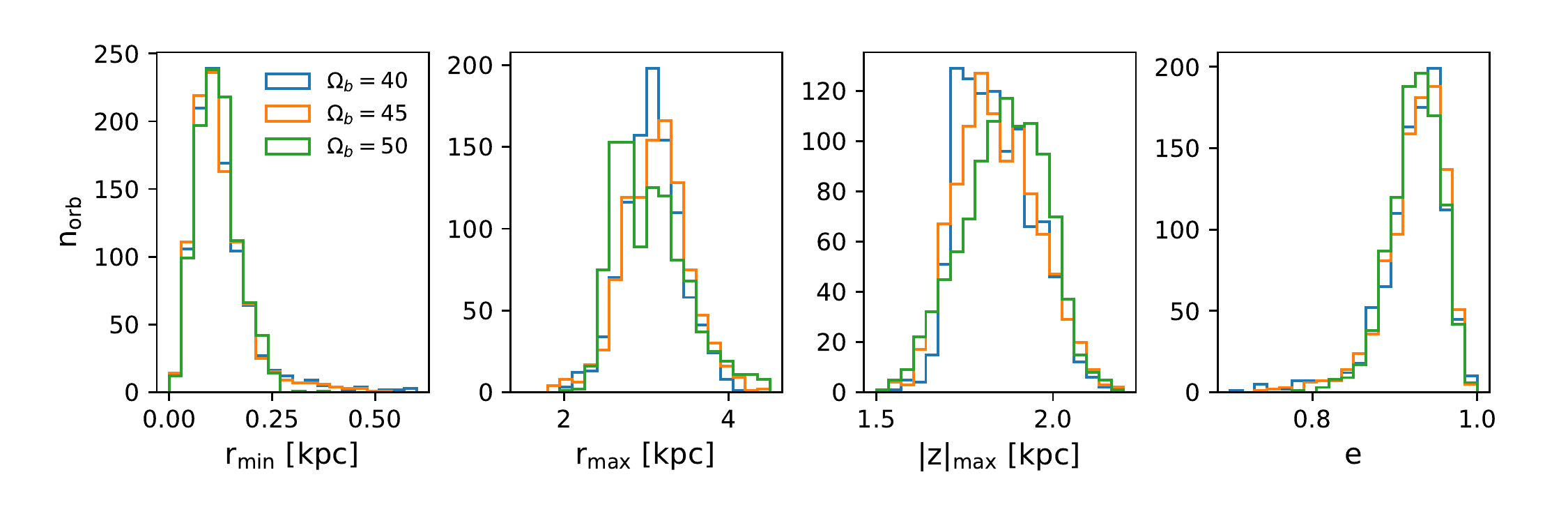}
    \caption{Distribution of orbital parameters for HP\,1, perigalactic distance $r_{\rm min}$, apogalactic distance $r_{\rm max}$, maximum vertical excursion from the Galactic plane $|z|_{\rm max}$, and eccentricity. The colours show the different angular speed of the bar, $\Omega_b=40$ (blue), 45 (orange), and 50 (green) km s$^{-1}$ kpc$^{-1}$.    
    }   
    \label{hist_orbits}
\end{figure*}

Distributions for the perigalactic distance, apogalactic distance, maximum vertical height and the eccentricity are presented in Figure \ref{hist_orbits}, the different colours represent the angular velocities we investigated. The orbits of HP\,1 have radial excursions between $\sim 0.1$ kpc to $\sim 4.0$ kpc, with maximum vertical excursions from the Galactic plane between $\sim 1.5$ kpc to $\sim 2.1$ kpc, and eccentricities $e>0.8$. The variation of the angular velocity seems to have an insignificant effect on the orbits.
In Table \ref{tab:Orb_para}, we present average orbital parameters of the set of orbits for HP\,1, the errors provide in each column are the standard deviation of the distribution. \citet{Baumgardt+18} also used Gaia DR2 PMs to integrate the orbit of HP 1 and calculate their perigalactic and apogalactic diactances, 0.56 kpc and 1.95 kpc, respectively. The differences with our orbital calculations are possibly due to the facts that they integrate backward for 2 Gyr only, in an axisymmetric Galactic potential, and the heliocentric distance is $\sim 200$ pc farther than our distance determination.

The orbital characteristics of HP\,1 are consistent with the bulge GCs that have [Fe/H]$\sim -1.0$ \citep{Perez-Villegas+18}, being well confined to the Galactic volume where bulge GCs are located, $R_{\rm GC} \leq 3.0$ kpc \citep{Bica+2016}.


\begin{table*}
\begin{center}
\caption{Monte Carlo average orbital parameters of HP\,1.}
\begin{tabular}{@{}ccccc@{}} 
\hline
$\Omega_b$ & $\langle r_{\rm min}\rangle$ & $\langle r_{\rm max} \rangle$ & $\langle |z|_{\rm max}\rangle $ &$\langle e\rangle$ \\
(km s$^{-1}$ kpc$^{-1}$) & (kpc) & (kpc) & (kpc) & \\
\hline
 40  & $0.128 \pm 0.079$ & $3.076 \pm 0.350$ & $1.841 \pm 0.104$ & $0.922 \pm 0.040$\\ 
45   & $0.124 \pm 0.070$ & $3.125 \pm 0.404$ & $1.843 \pm 0.114$ & $0.924 \pm 0.037$  \\
50   & $0.119\pm 0.050$ & $3.064 \pm0.460$ & $1.861\pm 0.116$ & $0.925 \pm0.031$ \\
\hline
\end{tabular}
\end{center}
\label{tab:Orb_para}
\end{table*}

\section{Summary and Conclusions}
\label{conclusions}

We obtained unprecedented deep NIR photometry in the crowded field of the moderately metal-poor  ([Fe/H]$=-1.06\pm0.10$) and $\alpha$-enhanced ([$\alpha$/Fe]$\sim+0.3$) inner bulge globular cluster HP\,1.
The GSAOI+GeMS detector at the Gemini-South telescope provides an unique opportunity to obtain MCAO images in $J$ and $K_{\rm S}$ filters. 
By combining our GSAOI data with archival \textit{HST} exposures, we computed the relative PMs of the stars in the core of HP\,1 and infer their membership to the cluster. This way, we created NIR and optical-NIR CMDs down to $K_{\rm S} \sim 19.5$ with only likely cluster members to analyze.

A statistical isochrone fitting technique following a Bayesian approach was employed to determine ages, distances and reddening in a self-consistent way. 
For this purpose, we used DSED and BaSTI $\alpha$-enhanced isochrones with canonical helium abundances ($Y\sim0.25$). 
Furthermore, a sample of 11 RR Lyrae stars (from the OGLE CVS, \citealt{Soszynski+14}, but also identified in the VVV survey, \citealt{Saito+12}) 
and a colour-colour diagram ($m_{\rm F606W} - K_{\rm S}$ vs. $J-K_{\rm S}$) were used to provide independent constraints for the distance, reddening and helium content. 
The main results from the isochrone fits can be summarized as follow: 
\begin{itemize}
\item The NIR and optical-NIR CMDs using DSED and BaSTI models confirm that HP\,1 is a fossil relic with an age of 12.8$^{+0.9}_{-0.8}$ Gyr.
\item The heliocentric and Galactocentric distances for this cluster are $6.59^{+0.17}_{-0.15}$ kpc and $1.7\pm0.3$ kpc, respectively.
\item The apparent distance moduli in $K_{\rm S}$ and $V$ are in very good agreement with the ones from the RR Lyrae stars using the recent $PM_{\rm K_{\rm S}}Z$ and $M_{V}-$[Fe/H] calibrations provided by the the \citet{Gaia17}.
\item A primordial He content is imposed by the mean magnitude of the RR Lyrae stars, otherwise the heliocentric cluster distance should be unrealistically high.
\item The E($B-V$) of HP\,1 is $1.15\pm0.02$, with good concordance between all isochrone fits. 
\item Even assuming uniform prior probability distributions, the recovered metallicity values are in very good agreement with the one from high resolution spectroscopy \citep{Barbuy+16}, adding reliability to the aforementioned results.
 \end{itemize}




We also performed an orbital analysis for HP\,1, taking into account the uncertainties of its parameters, through a Monte Carlo method. We found that the HP\,1's orbit is typical for moderately metal-poor bulge GCs,
with apogalactic distances not farther than $\sim 4$ kpc. These orbits present an average apogalactic distance $\langle r_{\rm max} \rangle  = 3.1 \pm 0.4$ kpc, which can be considered as a representative distance where it is possible to find fossil relics of the early fast chemical enrichment in the Galactic bulge. 

To establish a comprehensive scenario about the formation of the MW bulge, an homogeneous statistical analysis applied to a large sample of bulge GCs with deep NIR proper-motion-cleaned CMDs is highly recommended. 

\section*{Acknowledgements}

We are thankful to the anonymous referee for a careful reading and useful suggestions. LK, BB   and EB acknowledge partial financial support from FAPESP, 
CNPq, and CAPES - Finance Code 001. RAPO acknowledges a CAPES master fellowship and
SOS a FAPESP IC fellowship 2016/20566-7. ML, SO, and DN acknowledge
partial support by the Dipartimento di Fisica e Astronomia dell'Universit\`a di
Padova. APV acknowledges FAPESP for the postdoctoral fellowship 2017/15893-1.

\appendix

\section{Geometric distortion}\label{GD}

The GD of GSAOI has been solved by different authors, e.g.,
\citet{2016SPIE.9909E..1GM} and \citet{2016ApJ...833..111D}, with
different techniques. In this paper, we followed the prescriptions
given by \citet{2003PASP..115..113A,AK04},
\citet{2009PASP..121.1419B,2010A&A...517A..34B},
\citet{2011PASP..123..622B} and
\citet{Libralato+14,Libralato+15}. Note that the
AO-induced changes of the mirror configuration are expected to make
the incoming light reaching the detector through a different path in
each image \citep[e.g.][]{2016SPIE.9909E..1GM}. As such, the GD is
expected to slightly change from image to image and our GD solution is
meant to be an average GD solution.

The data set employed to solve for the GD of GSAOI is that of the GC
NGC\,6624 (program GS-2013A-Q23, PI: D. Geisler), which was also
adopted by \citet{2016ApJ...833..111D} for the same GD-correction
task. This data set was obtained when the GC was close to the zenith,
with optimal and stable seeing conditions. The dither pattern of these
observations is not ideal for a GD self-calibration \citep[see
  discussion in][]{Libralato+14}, but it is large enough to
allow us to calibrate the GD on to an external, distortion-free
reference frame.

As \citet{2016ApJ...833..111D}, we chose a \textit{HST} catalog to use
as reference. We adopted the
public-available\footnote{\href{http://groups.dfa.unipd.it/ESPG/treasury.php}{http://groups.dfa.unipd.it/ESPG/treasury.php}},
preliminary-release catalog of the ``\textit{Hubble Space Telescope}
UV Legacy Survey of Galactic GCs'' \citep[GO-13297, PI: G.
  Piotto]{Piotto+15}. Stellar positions are those listed in
the ACS Globular Cluster Treasury Survey catalog \citep[GO-10775, PI:
  A. Sarajedini, see][]{2008AJ....135.2055A}.

In our GD analysis, we considered only stars members of NGC\,6624 to
avoid relative PMs (GSAOI images were obtained 7 years after \textit{HST} data)
to mask out the systematic GD trends with an amplitude lower than that
induced by the relative cluster-field motion. At this aim, we made
advantage of stellar displacements available in this release of the
catalog obtained by combining the ACS Treasury and the UV Legacy
Survey data.

The GD correction was performed independently for each chip by means
of a third-order polynomial. As in \citet{2003PASP..115..113A} and
\citet{2010A&A...517A..34B}, we adopted the center of each chip
$(x,y)$ $=$ $(1024,1024)$ as reference pixel with respect to which
solve for the GD.

First, we selected bright, unsaturated, well-measured (\texttt{QFIT}
$<$ 0.05) stars in each GSAOI catalog. Then, we transformed stellar
positions from the \textit{HST} reference frame (hereafter master
frame) on to the reference frame of each chip/image. We adopted
four-parameter linear transformations (rigid shifts in the two
coordinates, one rotation, and one change of scale) to bring the
stellar position from one frame to the other. Positional residuals
were computed as the difference between the distortion-free,
master-frame-transformed positions and the GSAOI positions.

For each chip, stellar positional residuals were collected into a $16
\times 16$ look-up table (grid elements of $128 \times 128$ GSAOI
pixel$^2$ each). We then computed the $3\sigma$-clipped average value
of the residuals in each cell. Finally, we calculated the coefficients
of the third-order polynomial with a linear least-square fit to the
average residuals.

GSAOI positions were corrected for GD by using the available
third-order polynomial correction. The entire procedure was iterated,
each time computing new, improved linear transformations between the
master frame and the GD-corrected GSAOI catalogs, and new positional
residuals to refine the coefficients of the third-order
polynomials. The iterations ceased when the difference between the
polynomial coefficients from one iteration to the previous one was
negligible.

\citet{2016ApJ...833..111D} solved for the GD of GSAOI detector with a
procedure similar to that described here. The main difference with the
work of \citeauthor{2016ApJ...833..111D} is that we set the
coefficients $a_1$ and $a_2$ of their Eq.~3 equal to zero. As
discussed in, e.g., \citet{2003PASP..115..113A} and
\citet{2010A&A...517A..34B}, this convention constraints the GD
solution to have at the center of the chip (i) the $x$ scale equal to
that of the chip at the position of the reference pixel, and (ii) the
$y$ axis in the GD-corrected frame aligned to its $y$ axis (always at
the position of the reference pixel). The coefficients $b_1$ and $b_2$
of their Eq.~4 were left free to assume any value from the fit because
the scale and the perpendicularity of the two axes can be different.

\begin{figure*}
  \includegraphics[width=\columnwidth,angle=0]{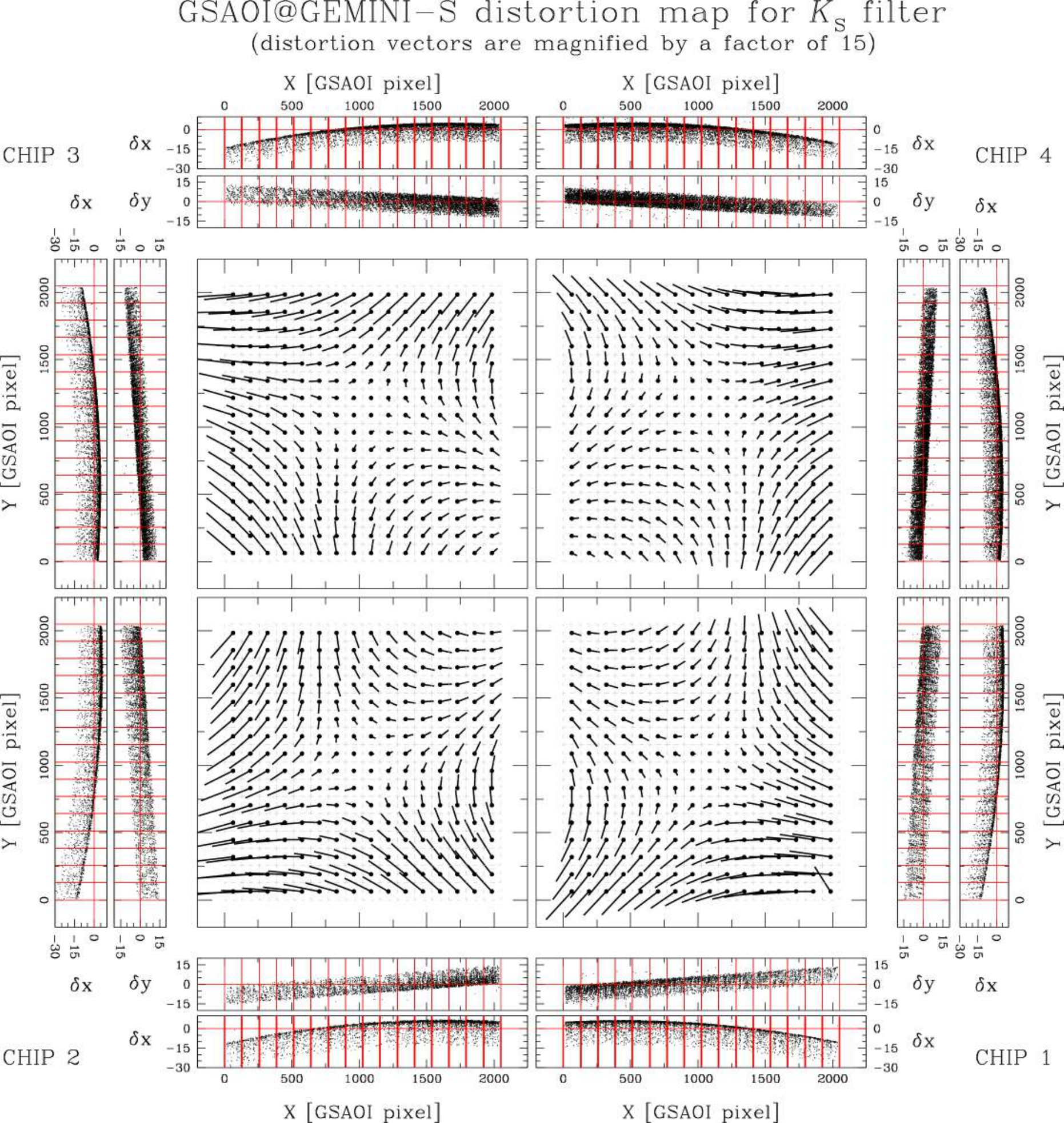}
  \includegraphics[width=\columnwidth,angle=0]{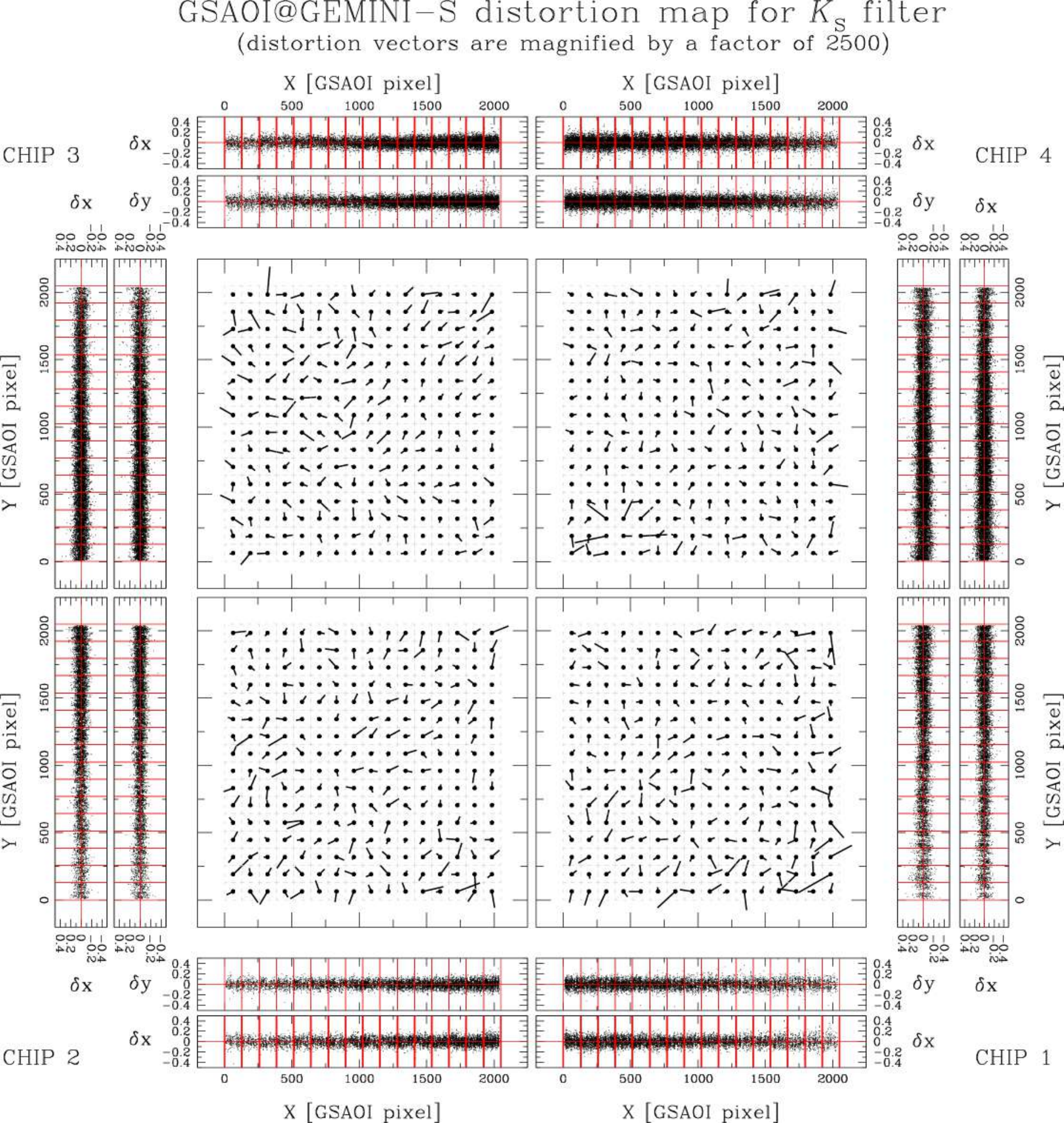}
  \caption{GSAOI $K_{\rm S}$-filter distortion maps before (left-hand
    panels) and after (right-hand panel) applying the GD
    correction. In the panels outside the distortion maps, we present
    the $\delta x$/$\delta y$ positional residuals (in GSAOI pixel) as
    a function of the $x$ and $y$ positions.}
  \label{mapgdKs}
\end{figure*}

\begin{figure*}
  \includegraphics[width=\columnwidth,angle=0]{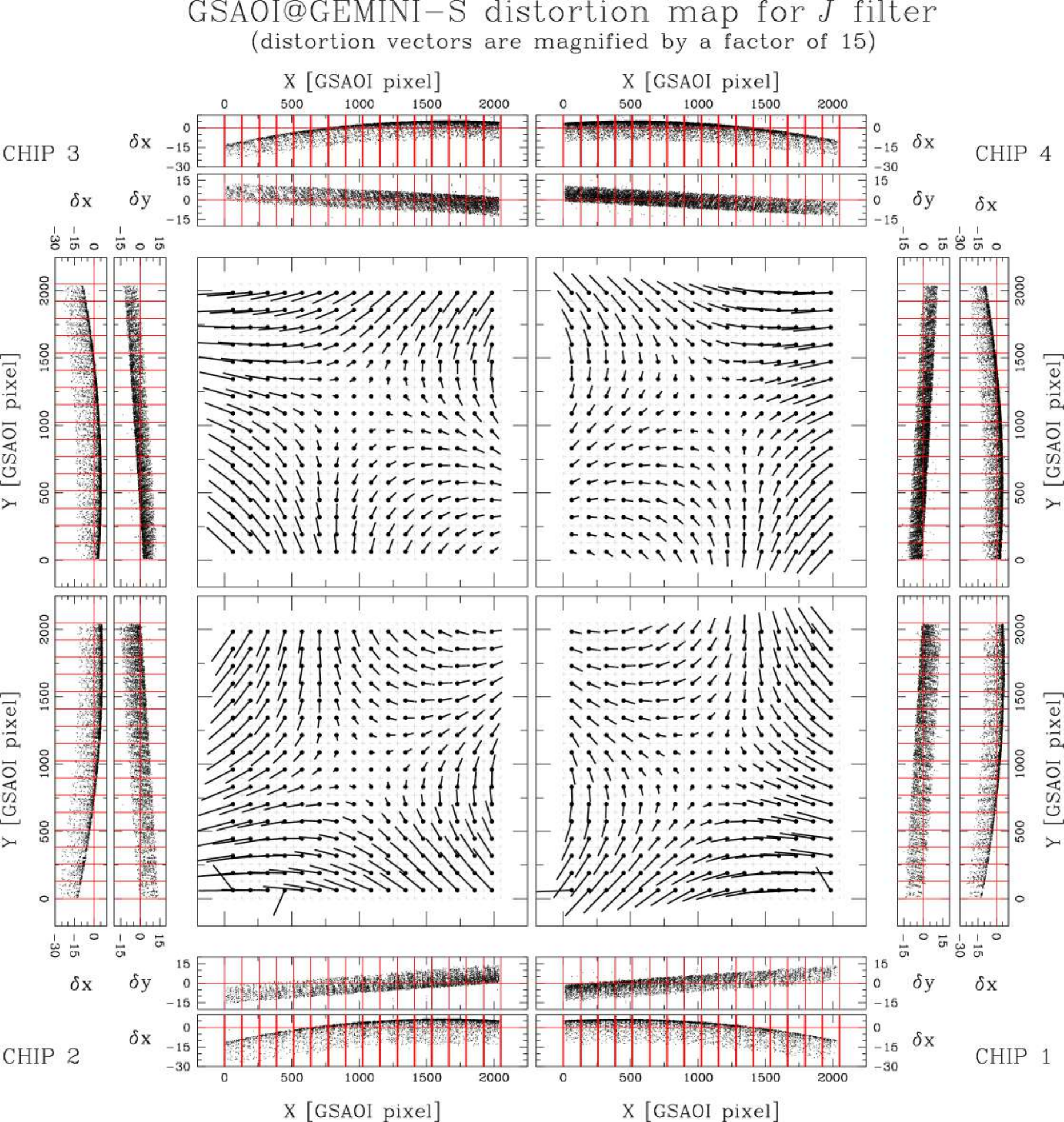}
  \includegraphics[width=\columnwidth,angle=0]{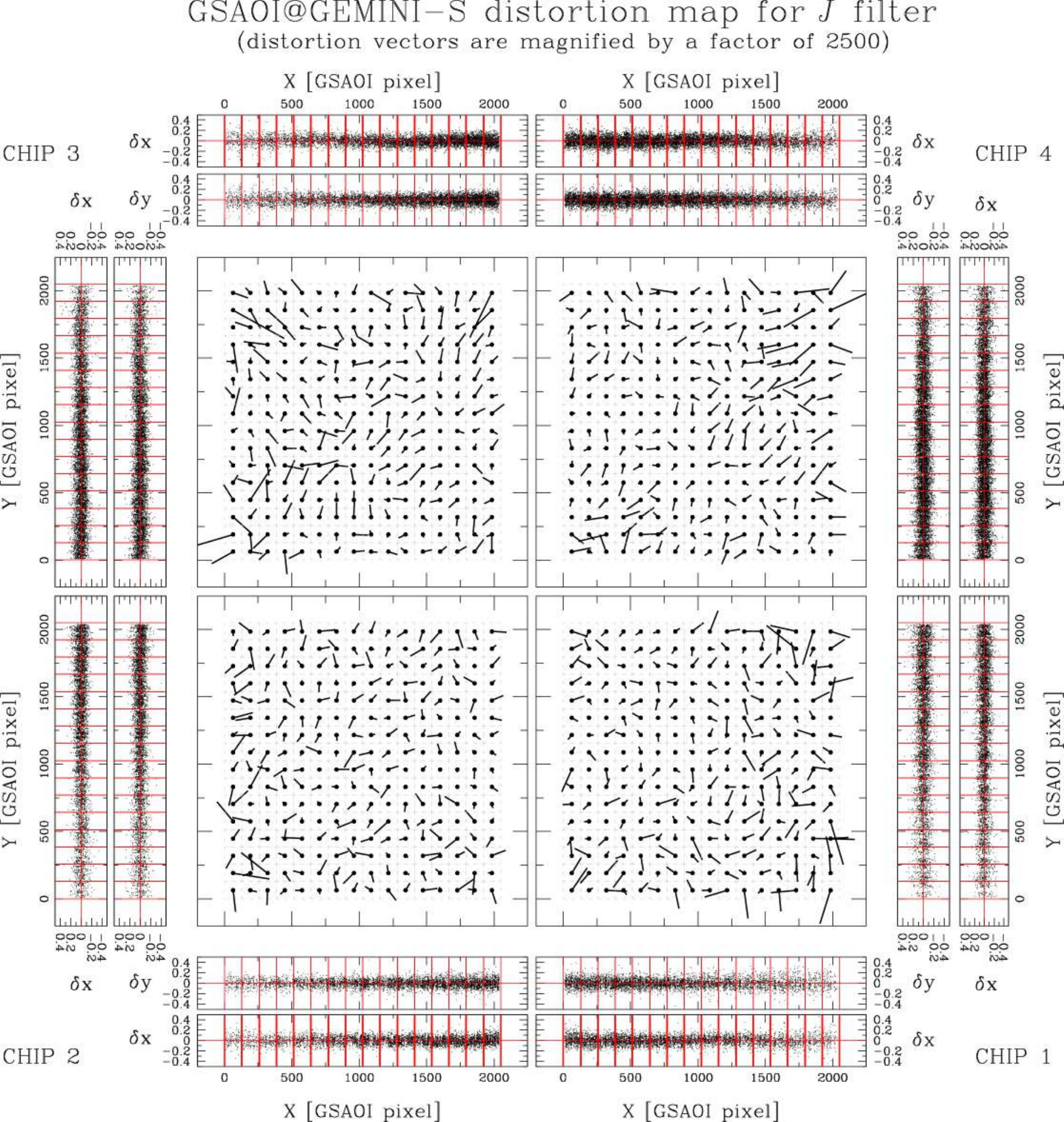}
  \caption{Same as in Figure~\ref{mapgdJ} but for the $J$ filter data.}
  \label{mapgdJ}
\end{figure*}

\begin{figure*}
  \includegraphics[width=\columnwidth,angle=0]{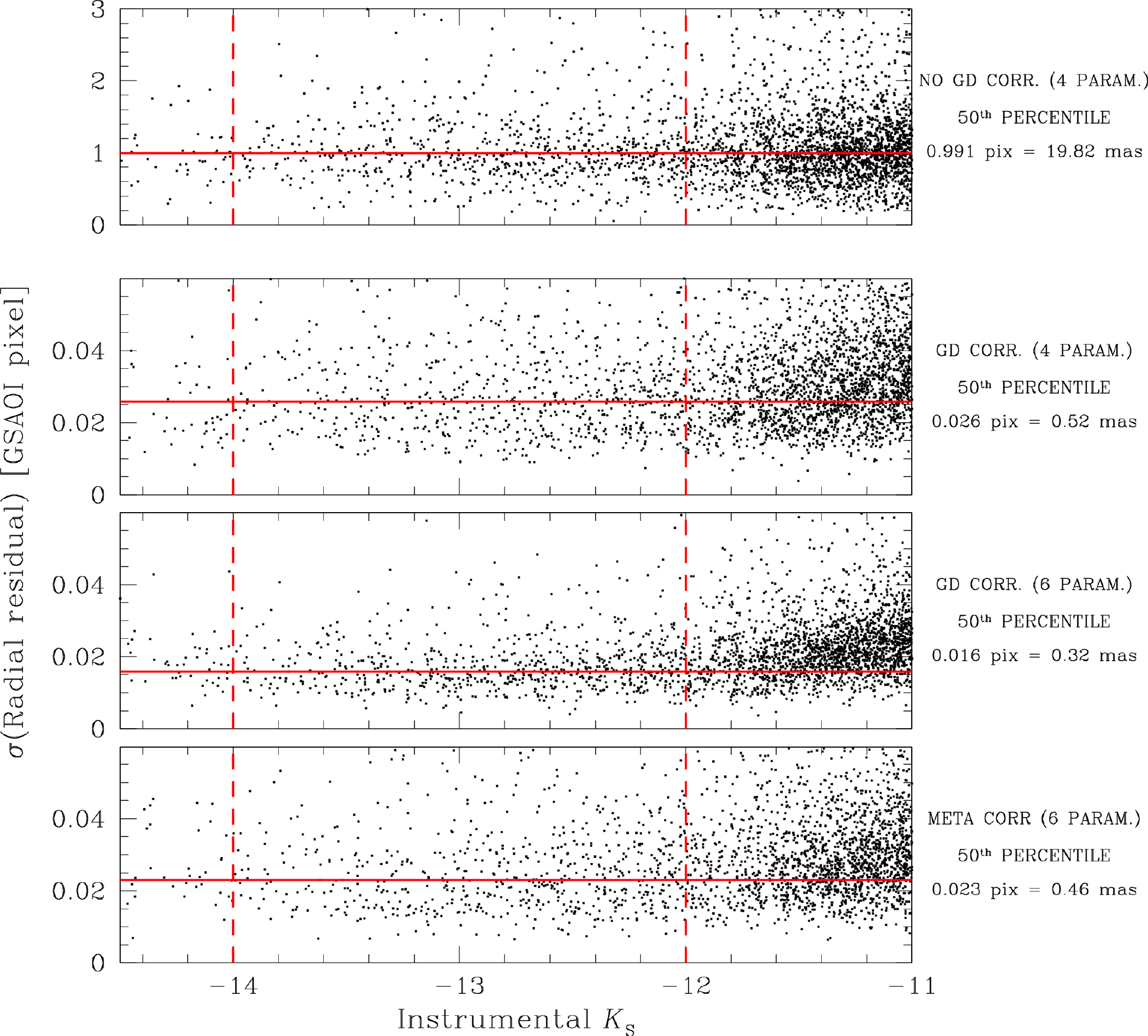}
  \includegraphics[width=\columnwidth,angle=0]{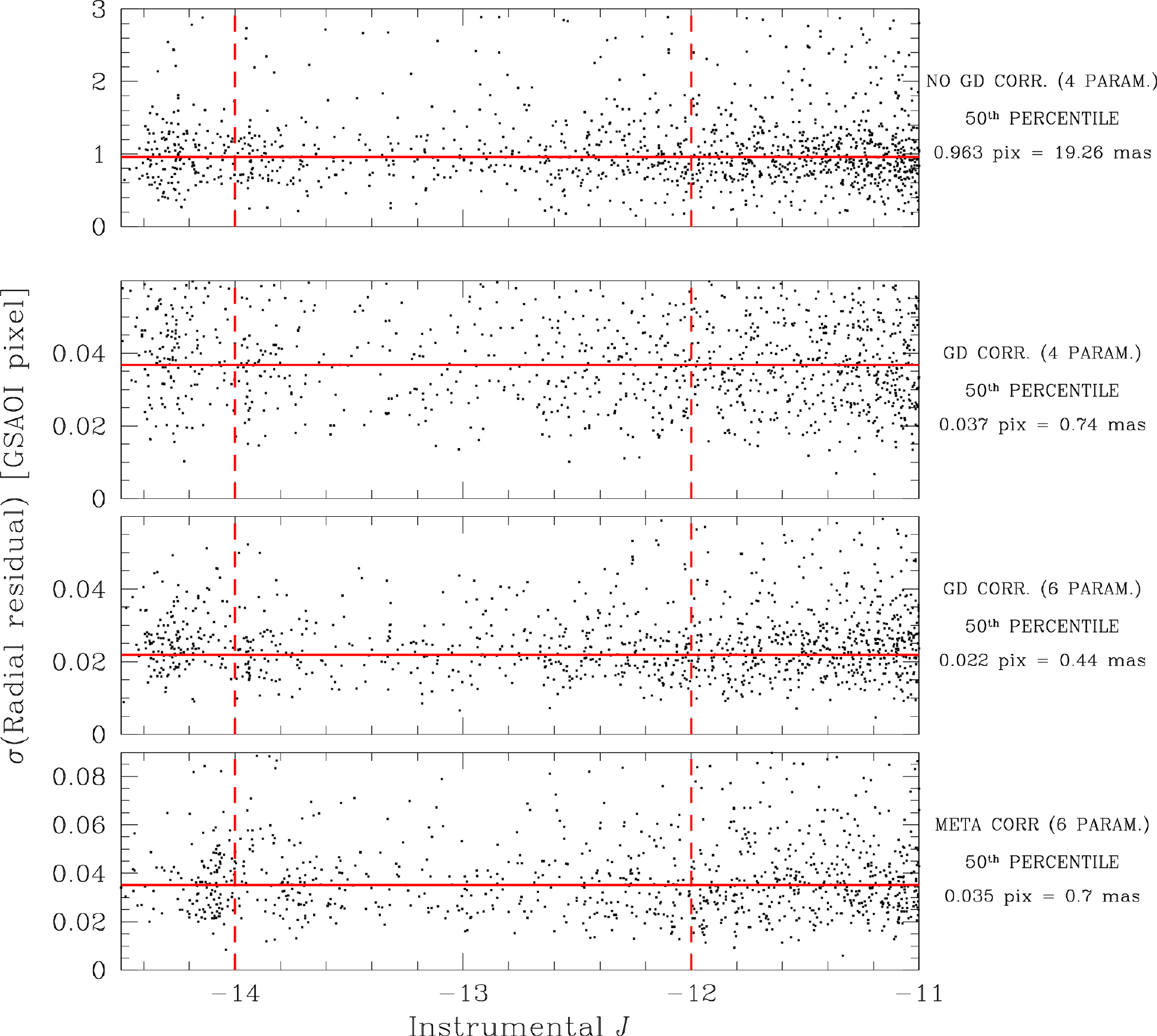}
  \caption{$\sigma$(Radial residuals) as a function of $K_{\rm S}$
    (left panels) and $J$ (right panels) instrumental magnitudes after
    each step of the GD correction. The red, solid horizontal line is
    set at the median value of $\sigma$(Radial residuals) for bright,
    unsaturated stars (between the two red, dashed vertical lines). In
    the bottom panels, we present the result obtained by adopting the
    meta-catalog solution.}
  \label{rmsgd}
\end{figure*}

The $K_{\rm S}$- and $J$-filter distortion maps before and after the
GD correction are shown in Figure~\ref{mapgdKs} and Figure~\ref{mapgdJ},
respectively. In Figure~\ref{rmsgd}, we present the $\sigma$(Radial
residuals) as a function of instrumental magnitude for $K_{\rm S}$-
(left) and $J$-filter (right) data after each step of the GD
correction. The $\sigma$(Radial residuals), computed as described in
\citet{Libralato+14,Libralato+15}, is indicative of the
accuracy of our GD solution. This value is defined as the rms of the
difference between the master-frame positions and the GD-corrected
stellar positions as measured in a given image transformed on to the
master-frame system. The master frame used for this computation is not
the \textit{HST} catalog but it was obtained by cross-identifying the
same stars in the GSAOI catalogs. We find that for bright,
well-measured stars, the median value of the $\sigma$(Radial
residuals) in the $K_{\rm S}$-filter data improves from $\sim 19.8$
mas to 0.52 mas (0.026 GSAOI pixel) in each coordinate. When we use
six-parameter linear transformations (that absorb most of the
residuals due to the variation of the telescope$+$optics system) to
transform stellar positions from the reference frame of a given
catalog on to the master frame, the value of $\sigma$(Radial
residuals) further improves to 0.32 mas. For $J$-filter data, we
measure instead a median $\sigma$(Radial residuals) of $\sim 0.44$
mas.

We also put the four GSAOI chips into a common, distortion-free
reference frame, hereafter meta catalog, with the same procedure
described in \citet{2010A&A...517A..34B}. In a nutshell, we
transformed the GD-corrected positions of each chip $k$ into the
GD-corrected system of chip 2 by means of four-parameter linear
transformations. The relations between the positions of a star in the
chip-$k$ system $(x_{\it k}^{\rm corr},y_{\it k}^{\rm corr})$ and that
in the chip-2 system $(x_{2}^{\rm corr},y_{2}^{\rm corr})$ are the
following:
\begin{equation*}
\tiny
  \begin{array}{ll}
  \left(
  \begin{array}{c}
    x_{2}^{\rm corr} \\
    y_{2}^{\rm corr} \\
  \end{array}
  \right) & 
  = \frac{\alpha_k}{\alpha_{2}}
  \left[
    \begin{array}{rr}
      \cos(\theta_k-\theta_{2}) & -\sin(\theta_k-\theta_{2}) \\
      \sin(\theta_k-\theta_{2}) &  \cos(\theta_k-\theta_{2}) \\
    \end{array}
    \right]
  \left(
  \begin{array}{c}
    x_{\it k}^{\rm corr}-1024 \\
    y_{\it k}^{\rm corr}-1024 \\
  \end{array}
  \right)
  \\
  \\ & +
  \left(
  \begin{array}{c}
    (x_2^{\rm corr})_{\it k} \\
    (y_2^{\rm corr})_{\it k} \\
  \end{array}
  \right)
  \end{array}\,.
\end{equation*}
The scale factor is indicated as $\alpha_{\it k}$, the orientation
angle as $\theta_{\it k}$, and the position of the center of chip $k$
in the reference frame of chip 2 as $(x_2^{\rm corr})_{\it k}$ and
$(y_2^{\rm corr})_{\it k}$, respectively. By construction, positions
in the chip-2 and in the meta systems are the same.

\begin{figure}
  \includegraphics[width=\columnwidth,angle=0]{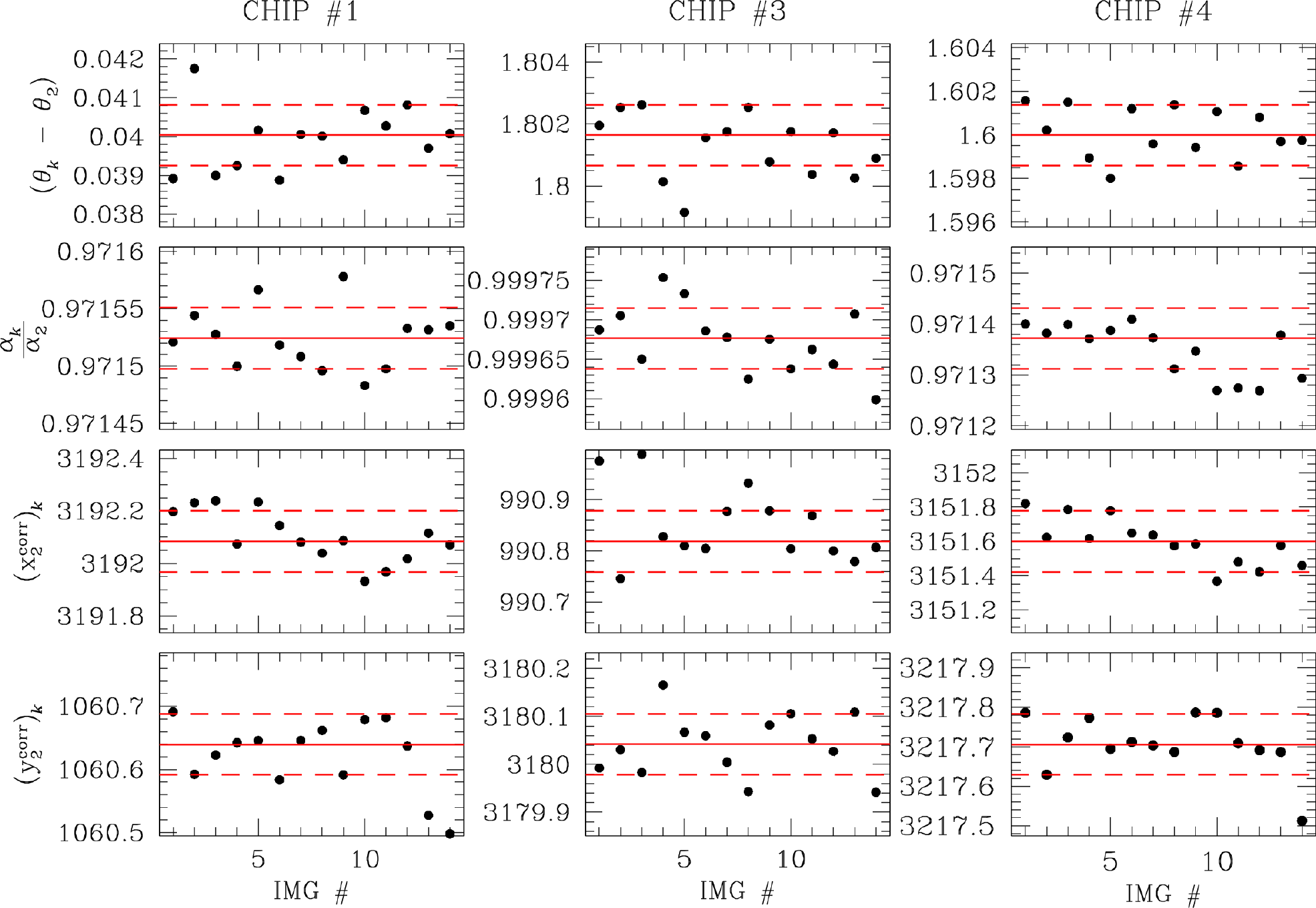}
  \caption{From top to bottom we present the relative angle (in
    degrees), scale, $x$ and $y$ positions (in GSAOI pixels) of the
    central pixel of chip 1 (left column), 3 (middle column) and 4
    (right column) with respect to chip 2 of the $K_{\rm S}$-filter
    images. The red, solid horizontal line is the median value of each
    parameter, the dashed lines are the $\pm1\sigma$ uncertainties.}
  \label{metaKs}
\end{figure}

The values of $\alpha_{\it k}$, $\theta_{\it k}$, $(x_2^{\rm
  corr})_{\it k}$ and $(y_2^{\rm corr})_{\it k}$ in each $K_{\rm
  S}$-filter image are shown in Figure~\ref{metaKs}. The final values
adopted for the $K_{\rm S}$-filter meta solution were computed as the
median values of these parameters (see Table~\ref{tabmetaKs}).

\begin{table*}
  \caption{Relative angle, scale and central-pixel positions of each
    chip relative to chip 2 obtained by using the $K_{\rm S}$-filter
    data.}
  \centering
  \label{tabmetaKs}
  \begin{tabular}{ccccc}
    \hline
    \hline
    Chip & $\theta_k-\theta_2$ & $\alpha_k/\alpha_2$ & $(x_2^{\rm corr})_k$ & $(y_2^{\rm corr})_k$\\
     & [deg] & & [GSAOI pixel] & [GSAOI pixel] \\
    \hline
    1 & $0.0400 \pm 0.0008$ & $0.97152 \pm 0.00003$ & $3192.08 \pm 0.12$ & $1060.64 \pm 0.05$ \\
    2 & 0.0000 & 1.00000 & 1024.00 & 1024.00 \\
    3 & $1.8016 \pm 0.0010$ & $0.99968 \pm 0.00004$ & $990.82 \pm 0.06$ & $3180.04 \pm 0.06$ \\
    4 & $1.6000 \pm 0.0014$ & $0.97137 \pm 0.00006$ & $3151.60 \pm 0.18$ & $3217.71 \pm 0.08$ \\
    \hline
  \end{tabular}
\end{table*}

Finally, we investigated how these parameters change by adopting a
different filter. We performed the same computation with the
$J$-filter data and the values are presented in
Table~\ref{tabmetaJ}. The most relevant difference is relative
scales. This discrepancy is related to the different optical path
introduced by the $J$ filter.

\begin{table*}
  \caption{As in Table~\ref{tabmetaKs} but computed by using the $J$-filter images.}
  \centering
  \label{tabmetaJ}
  \begin{tabular}{ccccc}
    \hline
    \hline
    Chip & $\theta_k-\theta_2$ & $\alpha_k/\alpha_2$ & $(x_2^{\rm corr})_k$ & $(y_2^{\rm corr})_k$\\
     & [deg] & & [GSAOI pixel] & [GSAOI pixel] \\
    \hline
    1 & $0.0435 \pm 0.0009$ & $0.97146 \pm 0.00004$ & $3192.00 \pm 0.11$ & $1060.66 \pm 0.13$ \\
    2 & 0.0000 & 1.00000 & 1024.00 & 1024.00 \\
    3 & $1.8043 \pm 0.0018$ & $0.99955 \pm 0.00003$ & $990.82 \pm 0.06$ & $3180.03 \pm 0.06$ \\
    4 & $1.6051 \pm 0.0027$ & $0.97137 \pm 0.00005$ & $3151.40 \pm 0.15$ & $3217.73 \pm 0.11$ \\
    \hline
  \end{tabular}
\end{table*}

The $\sigma$(Radial residuals) obtained by using the meta GD solution
is of the order of 0.46 and 0.7 mas for $K_{\rm S}$- and $J$-filter
data (bottom panels of Figure~\ref{rmsgd}), respectively. The accuracy
of the meta solution is slightly worse than the single-chip correction
because of the additional uncertainties in the linear terms of the
GD. Despite this worse accuracy, the distortion-free meta catalogs can
be useful when there are not enough stars to cross-identify all chips
(e.g., small-absent dither in the observations or sparse fields).

\bsp	

\label{lastpage}
\end{document}